\documentclass[12pt,eqsecnum,nofootinbib,floats,aps,prd,floatfix,titlepage,tightenlines]{revtex4} 

\usepackage{todonotes,soul}

\usepackage{graphicx}
\usepackage{graphics}
\usepackage{bm}
\usepackage{amssymb}
\usepackage{amsmath}
\usepackage{multirow}
\usepackage{hhline}
\usepackage{mathrsfs}

\usepackage[flushleft]{threeparttable}

\begin{document}
\title{Vacuum Quantum Stress Tensor Fluctuations:\protect\\ A Diagonalization Approach}

\author{Enrico D. Schiappacasse}
\email{Enrico.Schiappacasse@tufts.edu}
\author{Christopher J. Fewster}
\email{chris.fewster@york.ac.uk}
\author{L. H. Ford}
\email{ford@cosmos.phy.tufts.edu}
\affiliation{$^{\ast,\,\ddagger}$ Institute of Cosmology, Department of Physics and Astronomy,\\
Tufts University, Medford, MA 02155, USA
\vskip 0.01in
$^{\dagger}$ Department of Mathematics, University of York, Heslington, York YO10 5DD, United Kingdom
\vskip 0.5in}
\date{\today}

\begin{abstract}
Large vacuum fluctuations of a quantum stress tensor operator can be described by the asymptotic behavior of the probability distribution of the time or spacetime averaged 
operator. Here we focus on the case of stress tensor operators averaged with a sampling function in time. The Minkowski vacuum state is not an eigenstate of the time-averaged 
operator, but can be expanded in terms of its eigenstates. We calculate the probability distribution and the cumulative probability distribution for obtaining a given value in a 
measurement of the time-averaged operator taken in the vacuum state. In these calculations, we use the normal ordered square of the time derivative of a massless scalar field in 
Minkowski spacetime as an example of a stress tensor operator. We analyze the rate of decrease of the tail of the probability distribution for different temporal sampling functions, 
such as compactly supported functions and the Lorentzian function. We find that the tails decrease relatively slowly, as exponentials of fractional powers, in agreement with previous 
work using the moments of the distribution. Our results lead additional support to the conclusion that large vacuum stress tensor fluctuations are more probable than large thermal 
fluctuations, and may have observable effects.
\end{abstract}

\maketitle

\baselineskip=19pt

\section{Introduction}

The definition and the use of the expectation value of a quantum stress tensor operator have been a topic of
intense study in recent decades. The semiclassical theory for gravity uses the renormalized expectation value of the
quantum matter stress tensor to give an approximate description of the effects of quantum matter fields on the gravitational field. As in the semiclassical theory of 
electromagnetic radiation, it is expected that this theory is a reasonable approximation to a more complete quantum theory of gravity coupled to matter fields. It is known that
a renormalized stress energy operator for quantum fields in curved spacetime is associated with quantum corrections to Einstein's equations, via higher order derivative 
terms~\cite{Horowitz:1978fq}. These corrections lead to physical effects, such as small scale factor oscillations around an expanding background universe and quantum 
particle creation~\cite{Schiappacasse:2016nei}. Moreover, this theory has been successful about giving a plausible description of the back reaction to black hole evaporation through 
Hawking radiation~\cite{ParkerDavid2009}. However, the semiclassical theory does not consider the quantum fluctuations of the stress tensor around its expectation value and 
their possible effects.
Several authors have studied a variety of physical effects associated with quantum stress tensor fluctuations~\cite{BorgmanFord2004andothers}. These effects include, 
for example, potentially observable gravity waves from quantum stress tensor fluctuations in inflationary models~\cite{Wu:2011gk}, effects of vacuum electric field fluctuations 
on light propagation in nonlinear materials~\cite{Bessa:2014pya, Bessa:2016uqb}, and barrier penetration of charged or polarizable particles through large vacuum radiation 
pressure fluctuations~\cite{Huang:2015lea, Huang:2016kmx}. 

In general, the physical effects of large fluctuations of a quantum stress tensor operator can be studied through 
the analysis of the probability distribution for the time or spacetime averaged operator. This probability distribution can be inferred (at least qualitatively) from 
the moments of the averaged operator,
and the exact distribution was found in a two-dimensional model in Ref.~\cite{Fewster:2010mc}.  The moments method was used in Ref.~\cite{Fewster:2012ej} 
to infer the probability distribution for several normal-ordered quadratic operators in four dimensional Minkowski spacetime with  Lorentzian time  averaging. These included
the square of the electric field and the energy densities of a massless scalar field and of the electromagnetic field.
This idea  was extended in Ref.~\cite{Fewster:2015hga} to compactly supported functions of time. These results predict an asymptotic form of the probability distribution function
 for large fluctuations of 
\begin{equation}
P(x) \sim c_0 x^b e^{-ax^c}\,,\,\,\,\,\,\,\,x\gg 1\,.
\label{asymptotic}
\end{equation}
Here the dimensionless variable $x$ is the measurement of the stress tensor fluctuations and $c_0, a, b, \text{ and } c$ are constants which depend on the sampling function. 
In the case of the Lorentzian time averaged electromagnetic energy density, for example, $a\sim 1$ and $c = 1/3$. Because thermal fluctuations are exponentially suppressed 
in energy, vacuum fluctuations can dominate over thermal fluctuations at large energies. However, the moments of a quantum stress tensor operator  grow very rapidly, to the extent that they might not uniquely
determine the probability distribution, so it is desirable to seek alternative methods.

In the present paper, we develop such an independent test of the moments approach for the study the probability distribution of time-averaged quantum stress tensor operators. 
The main idea is to diagonalize the time-averaged operator through a change of basis and calculate the cumulative probability distribution function of their quantum fluctuations 
in the vacuum state.
We are interested in checking the behavior predicted by the high moments approach, and in determining which modes and particle numbers give the dominant contribution 
to the large fluctuations. Unlike the moments approach, which primarily gives  information about the asymptotic behavior of the probability distribution for large vacuum stress
 tensor fluctuations, the diagonalization approach in principle gives a unique probability distribution for a broad range of fluctuations $x$. 
We take the normal ordered square of the time derivative of a massless scalar field in Minkowski spacetime as our stress tensor operator, and find the
tail of the probability distribution for different temporal sampling functions, specifically a class of compactly supported functions and the Lorentzian function. The tails
decrease relatively slowly, as exponentials of fractional powers, in agreement with previous results using the moments of the distribution.     

The paper is organized as follows: In Sec.~\ref{sec:moments}, we review the main results of Ref.~\cite{Fewster:2015hga} on the high moments approach to the analysis of the 
probability distribution for quantum stress tensor operators. In Sec.~\ref{sec:diagonal}, we develop an independent approach to the study of probability distributions based on the 
diagonalization of the operator. In Sec.~\ref{sec:masslessfield}, we show the numerical results obtained for different time sampling functions.  In Sec.~\ref{sec:summary}, we 
summarize and discuss the main results of the paper.   
   
\section{Moment-based approach to the probability distribution} \label{sec:moments}
 
Here we review the main results of Ref.~\cite{Fewster:2015hga}. Working in $4$-dimensional Minkowski spacetime, let $T(t,{\bf{x}})$ be a operator which is a quadratic 
function of a free field operator and define its time average with a real-valued sampling function $f(t)$ by
\begin{equation}
\overline{T} = \int_{-\infty}^{\infty}\mathopen{:}T(t,{\bf{x}})\mathclose{:}f(t)dt\,.
\label{T}
\end{equation}
We will consider measurements of the 
 time average $\overline{T}$ rather than $T$. The sampling function has a characteristic width $\tau$ and should decay quickly as $|t| \gg \tau$. One example is a 
 Lorentzian function, used in Ref.~\cite{Fewster:2012ej}, whose mathematical expression and Fourier transform are given by
\begin{equation}
f_L(t) = \frac{\tau}{\pi(t^2+\tau^2)}\,\,\,\,\, \text{and} \,\,\,\,\,\hat{f}_L(\omega) = e^{-|\omega \tau|}\,,
\label{FTlorentzian} 
\end{equation}
where the Fourier transform of $f_L(t)$ and its normalization are given by
\begin{equation}
\hat{f}_L(\omega) = \int_{-\infty}^{\infty} dt\,e^{-i\omega t}f_L(t)\,\,\,\text{and}\,\,\,\hat{f}_L(0)=1\,.
\label{FT}
\end{equation}
However, if the measurement of the operator occurs in a finite interval of time, the sampling function is better described by a smooth and compactly supported function. This kind of 
sampling function is strictly zero outside a finite region, avoiding the long temporal tails of functions like the Lorentzian. It therefore gives a better description of a measurement which 
begins and ends at finite times. We will be interested in compactly supported nonnegative functions whose Fourier transform has the following asymptotic form when
$\omega \tau \gg 1$: 
\begin{equation}
\hat{f}(\omega) \sim \gamma\, e^{-\beta|\omega \tau|^{\alpha}}\,,
\label{FTCSFunction} 
\end{equation}    
where $\alpha$,  $\gamma$, and $\beta$ are constants. Here $\alpha \in (0,1)$ is a decay parameter which defines the rate of decrease of 
$\hat{f}(\omega)$ (values $\alpha\ge 1$ are incompatible with $f$ having compact support). 
It is worth emphasising that $\tau$ does not directly measure the support of $f$, but rather indicates the shortest characteristic timescale associated with $f$; 
in our examples, this will characterise the switch-on and switch-off regions.

For any given $f$ (compactly supported or not) define the $n$-th moment of the normal-ordered time-averaged quadratic operator $\overline{T} $, Eq.~(\ref{T}), as
\begin{equation}
\mu_n = \langle 0|(\overline{T}) ^n|0 \rangle\,,
\end{equation}
where $|0 \rangle$ is the Minkowski vacuum vector of the theory. As we will now see, 
the form of the Fourier transform $\hat{f}$ defines the rate of growth of the moments $\mu_n$ and, as a result, the probability for large fluctuations.

In the first instance, we work in a box of finite volume and express $\overline{T}$ in a mode sum of creation and annihilation bosonic operators as
\begin{equation}
\overline{T}  = \sum_{ij} \left( \tilde{A}_{ij}a^{\dagger}_i a_j + \tilde{B}_{ij}a_ia_j + \tilde{B}^{*}_{ij}a^{\dagger}_ia^{\dagger}_j \right)\,,
\label{Texpansion} 
\end{equation}  
where $\tilde{A}_{ij}$ and $\tilde{B}_{ij}$ are components of symmetric matrices $\tilde{A}$ and $\tilde{B}$, which have the functional forms
\begin{gather}
\tilde{A}_{ij} \propto (\omega_i \omega_j)^{1/2} \hat{f}(\omega_i - \omega_j)\,,\\
\tilde{B}_{ij} \propto (\omega_i \omega_j)^{1/2} \hat{f}(\omega_i + \omega_j)\,,
\end{gather} 
where $\omega_i$ are the mode frequencies. Precise forms of $\tilde{A}$ and $\tilde{B}$ will be given when we come to specific examples in Section~\ref{sec:masslessfield}. The 
moment $\mu_n$ can be expressed as an $n$-th degree polynomial in these components.
As $n$ increases, the number of terms in the expression for the $n$-th moment grows rapidly. Fortunately, 
only one term gives the dominant contribution for $n \gg 1$:
\begin{equation}
M_n = 4 \sum_{j_1\cdots j_n} \tilde{B}_{j_1 j_2}\tilde{A}_{j_2j_3}\tilde{A}_{j_3j_4}\cdots\tilde{A}_{j_{n-1}j_n}\tilde{B}^*_{j_nj_1}\,.
\end{equation}
First, $\tilde{B}_{j_1j_2}$ and $\tilde{B}^*_{j_nj_1}$ have to begin and end, respectively, the expression for $M_n$
because $\tilde{B}^*_{ij}a^{\dagger}_ia^{\dagger}_j$ and $\tilde{B}_{ij}a_ia_j$ in Eq.~(\ref{Texpansion}) are the only terms which do not annihilate the vacuum from the left and right, 
respectively. Second, all the remaining coefficients in $M_n$ are $\tilde{A}_{ij}$'s, which fall slower than $\tilde{B}_{ij}$'s when $\omega_i$ becomes large. This arises because the $
\tilde{A}_{ij}$ involve a difference in frequencies, as opposed to the sum in the $\tilde{B}_{ij}$.  Provided that $\hat{f}\ge 0$, all the terms contributing to the $n$-th moment are nonnegative, so $M_n$ is actually a lower bound 
on $\mu_n$, which will gives us a lower bound on the probability distribution for large vacuum fluctuations. 

To be more specific, now consider the time average of $\mathopen{:}\dot{\phi}^2\mathclose{:}$, where $\phi$ is a massless scalar field in four-dimensional Minkowski spacetime. 
Then, passing to a continuous mode sum, the dominant term takes the form
\begin{equation}
M_n = \frac{1}{(2\pi^2)^n} \int_0^{\infty} d\omega_1\cdots d\omega_n (\omega_1\cdots \omega_n)^3\hat{f}(\omega_1 + \omega_2)\hat{f}(\omega_2-\omega_3)\cdots \hat{f}
(\omega_{n-1}-\omega_n)\hat{f}(\omega_n+\omega_1)\,.
\end{equation}
If $\hat{f}$ has the asymptotic form~\eqref{FTCSFunction}, then the dominant term has the asymptotic form, in units in which $\tau = 1$,
\begin{equation}
{M_n \sim \frac{3!\gamma^2 [2\pi f(0)]^{n-2}\Gamma\left[ (3n+2)/\alpha - 4\right]}{(2\pi^2)^n\alpha^5(2\beta)^{(3n+2)/\alpha}}}
\label{Mnasymptotic}
\end{equation}
for $n \gg 1$,
where $f(0)= (2\pi)^{-1}\int_{-\infty}^{\infty}d\omega \hat{f}(\omega)$ (see Sec.~IV of~\cite{Fewster:2015hga}).  The most important part of this expression is the gamma function factor, which leads a rapid rate of growth of 
the high moments, $M_n \propto (3n/\alpha)!$. Thus, the parameter $\alpha$ is crucial in determining the rate of growth of the moments when $n \gg 1$. 

The goal is to use the asymptotic form for the moments, Eq.~(\ref{Mnasymptotic}), to obtain information about the probability distribution for large vacuum fluctuations. Return to 
arbitrary units for the characteristic timescale $\tau$.
Let $P(x)$ be the probability density for the distribution of the dimensionless variable 
$x = \overline{T}\tau^4$ in measurements of $\overline{T}$ in the vacuum state. While there is
no upper bound on the values of $x$ that can arise -- and therefore no upper bound on the support of $P$ -- there is a lower bound $x>-x_0$ for some $x_0>0$. There is a deep 
connection between this feature of the stress tensor probability distribution and quantum inequality bounds, which is explained in detail in 
Refs.~\cite{Fewster:2010mc, Fewster:2012ej}. We define the tail distribution (also called the complementary  cumulative distribution function), $P_{>}(x)$, as the probability 
of finding any value $y\ge x$ in a measurement 
\begin{equation}
P_{>}(x) = \int_{x}^{\infty} P(y)dy 
\label{Pcumulative}
\end{equation} 
and of course $P$ is normalized so that $P_{>}(x)=1$ for $x\le -x_0$. The $n$-th moment of $\overline{T}$ can be written in terms of $P$ as
\begin{equation}\label{momentinfunctionofPX}
\mu_n = \tau^{-4n} \int_{-x_0}^\infty x^n P(x)\,dx
\end{equation}
and this can be compared with the the asymptotic form of the dominant contribution $M_n$, Eq.~(\ref{Mnasymptotic}), to infer information about $P(x)$ and $P_{>}(x)$. 
In this way, we are led to consider the asymptotic forms 
\begin{equation}
P(x) \sim c_0x^be^{-ax^c}\,,\,\,\,\,\,\text{and}\,\,\,\,\,P_{>}(x)\sim 1 - \frac{c_0a^{-(1+b)/c}}{c}\Gamma\left( \frac{1+b}{c},ax^c \right)\,,
\label{asymptoticPX}
\end{equation} 
for large vacuum fluctuations, $x \gg 1$,
where $c_0, a, b,$ and $c$ are constants to be determined, and for which the corresponding moments obey 
\begin{equation}
\mu_n \approx c_0\int_{-x_0}^{\infty} x^{n+b}e^{-ax^c}dx=\frac{c_0}{c}a^{-(n+b+1)/c}\,\Gamma[(n+b+1)/c]\,.
\label{MnasymptoticPX}
\end{equation}
when $n$ becomes large. 
 The similarity between this expression and the asymptotic form for $M_n$, Eq.~(\ref{Mnasymptotic}), is evident, and leads to the identifications
  \begin{equation}\label{eq:identifications}
 c = \frac{\alpha}{3}\,,\,\,\,b=-\frac{(4\alpha+1)}{3}\,,\,\,\,a=2 {\beta}\left( \frac{f(0)}{\pi} \right)^{-\alpha/3}\,,\,\,\,c_0=
 ca^{(1+b)/c}\,3! {\gamma^2}\alpha^{-5}\,(2 {\beta})^{-2/\alpha}[2\pi f(0)]^{-2}\,.
 \end{equation}
 However, the situation is a little bit more subtle, because it is not guaranteed that a set of moments
 growing as fast as $(3n/\alpha)!$ (for $\alpha<1$) determines a unique probability distribution~\cite{Simon1998}. Fortunately, the difference between two probability 
 distributions with the same moments is just an oscillatory function, which does not add any interesting feature to the general form of $P(x)$ for our purposes. 
 Therefore the parameters in 
 {Eq}.~\eqref{eq:identifications} should provide  a good approximation to the asymptotic behavior of $P(x)$ and $P_>(x)$.  Rigorous arguments to this effect are given in Sec.~VI of~\cite{Fewster:2012ej}.
 
 The argument just given applies to the case of a compactly supported function with asymptotics given by   Eq.~(\ref{FTCSFunction}). For the case of a non-compactly 
 supported sampling function such as a Lorentzian, Eq.~(\ref{FTlorentzian}), a slightly different argument is needed to compute the asymptotic form of the
 dominant contribution $M_n$, as is explained in detail in Ref.~\cite{Fewster:2012ej}. 
 However, the analysis of high moments still leads to an asymptotic form for $P(x)$ given by Eq.~(\ref{asymptoticPX}) with $c=1/3$. This is consistent with the $\alpha\to 1$ limit of 
 the relation $c = \alpha/3$ derived for compactly supported functions, in which limit the asymptotic form~\eqref{FTCSFunction} 
 agrees with that of the Lorentzian~\eqref{FTlorentzian}, with $\gamma = \beta = 1$. 
 
 In general, we see that the decay parameter $\alpha$ in the asymptotic form of the sampling function's Fourier transform determines the rate of decay in $P(x)$ for large $x$, and 
 hence the probability of large   vacuum fluctuations. The smaller $\alpha$ is, the more slowly the tail decreases and the greater the probability of large fluctuations becomes. 
 For compactly supported functions, the value of $\alpha$  is related to the rate of switch-on and switch-off of $f(t)$. [See Eqs.~(51) and (52) in Ref.~\cite{Fewster:2015hga}.] 
 
\section{Diagonalization of the quadratic bosonic stress tensor} \label{sec:diagonal}

So far, we have studied the probability distribution for quantum stress operators by analyzing the behavior of high moments of these operators. Now we proceed to 
develop an independent test of the moment-based approach, in which we diagonalize $\overline{T}$ and express the Minkowski vacuum vector in the basis of its eigenstates.
Note that the vacuum is not in general an eigenstate of the time averaged quantum stress tensor operator, $\overline{T}$; indeed, this would be incompatible with the Reeh--Schlieder 
theorem if the sampling
function is compactly supported. Using the expression for the vacuum in terms of the new basis allows us to calculate the probability distribution function of obtaining  a 
specific result in a measurement of $\overline{T}$. This approach can yield information about the contribution of various modes and occupation numbers to the 
probability distribution, in addition to providing a uniquely defined probability distribution.

\subsection{Bogoliubov diagonalization}
\label{diagonalization}
We express a general quadratic operator $H$ as a mode sum involving bosonic creation and annihilation operators  for $N$ modes as
\begin{equation}
H = \frac{1}{2} \sum_{ij}^N \left( a_i^{\dagger}D_{1ij}a_{j} + a_i^{\dagger}D_{2ij}a_j^{\dagger}+a_iD_{3ij}a_j+a_iD_{4ij}a_j^{\dagger}  \right)\,,
\label{TColpa}
\end{equation}
where 
\begin{equation}
[a_i,a_j^{\dagger}]=\delta_{ij}\openone\,\,\,\,\text{and}\,\,\,\,[a_i,a_j]=[a_i^{\dagger},a_j^{\dagger}]=0\,,
\label{commutation}
\end{equation}
and $\openone$ is the identity operator. 
Here the coefficients of Eq.~(\ref{TColpa}) correspond to elements of $N$-square matrices $\lbrace {D_r} \rbrace _{r=1}^{4}$ which form the so-called dynamical matrix 
\begin{equation}
\mathcal{D} = \left( \begin{array}{cc}
D_1 & D_2 \\
D_3 & D_4 \end{array} \right)\,.  
\end{equation}
Here we follow an approach developed by Colpa~\cite{Colpa1978} for the diagonalization of $\mathcal{D}$. This approach was previously applied to stress tensor operators by 
Dawson~\cite{Dawson2006}, who was primarily concerned with quantum inequality bounds on expectation values. 
The diagonalization of the quadratic operator $H$ implies a homogeneous linear transformation (Bogoliubov transformation~\cite{Bogoliubov1947}) to go from the original set of 
bosonic operators, $(a_i,a_i^{\dagger})_{i=1}^N$, to a new
one, $(b_i,b_i^{\dagger})_{i=1}^N$, in which $H$ takes a diagonal form. For our purposes, we consider the case $D_1=D_4=F$ and $D_2=D_3=G$ with
$F$ and $G$ real and symmetric matrices. Under these conditions, we may normal order the operator $H$ in Eq.~(\ref{TColpa}) to obtain 
\begin{equation}
\mathopen{:}H\mathclose{:} = \frac{1}{2}\left( 2{\bf{a}}^{\dagger}F{\bf{a}} + {\bf{a}}^TG{\bf{a}}+{\bf{a}}^{\dagger}G{\bf{a}}^{\dagger\,T}\right)\,,\,\,\,\,\,\text{with}\,\,\,\,\,
{\bf{a}}\equiv  \left( \begin{array}{c}
a_1 \\
a_2\\
\vdots \\
a_N \end{array}\right)\,\,\,\,\text{and}\,\,\,\,{\bf{a}}^{\dagger}\equiv   \begin{pmatrix}
 a_1^{\dagger} & a_2^{\dagger} & \cdots & a_N^{\dagger} \end{pmatrix} \,, 
\label{Trewritten}
\end{equation}
and the superscript $T$ denotes a transpose. 
Here we have combined the first and last terms in Eq.~(\ref{TColpa}) using the fact that $F$ is real and symmetric. Note that the operator $\overline{T}$ in Eq.~(\ref{Texpansion}) takes 
this form, in the case of infinite $N$, where $F = \tilde{A}$ and $G = 2\tilde{B}$. An important observation is that we may use the canonical commutation relations \eqref{commutation} to write
\begin{equation}
\mathopen{:}H\mathclose{:} =  \frac{1}{2}\, \begin{pmatrix} {\bf{a}}^\dagger & {\bf{a}}^T\end{pmatrix}\begin{pmatrix} F & G \\ G & F\end{pmatrix}
\begin{pmatrix} {\bf{a}} \\ {\bf{a}}^{\dagger T}\end{pmatrix} - \frac{1}{2}\text{Tr}(F) \openone\,.
\label{eq:H-matrixform}
\end{equation} 
Now we apply a Bogoliubov transformation 
\begin{equation}
{\bf{a}}=A{\bf{b}}+B{\bf{b}}^{\dagger\,T}\,,\,\,\,\,\,\text{with}\,\,\,\,\,
{\bf{b}}\equiv  \left( \begin{array}{c}
b_1 \\
b_2\\
\vdots \\
b_N \end{array}\right)\,\,\,\,\text{and}\,\,\,\,{\bf{b}}^{\dagger}\equiv  \begin{pmatrix} b_1^{\dagger} &  b_2^{\dagger} & \cdots &  b_N^{\dagger} \end{pmatrix}\,,
\label{Bogoliubovtransformation}
\end{equation} 
where $A$ and $B$ are real $N \times N$ matrices, and the new set of bosonic operators satisfy the usual commutation relations 
$[b_i,b_j^{\dagger}]=\delta_{ij}\openone\,\text{and}\,[b_i,b_j]=[b_i^{\dagger},b_j^{\dagger}]=0$. 
Note that the commutation relations for the $a$ and $a^\dagger$ operators and the Bogoliubov transformation, Eq.~(\ref{Bogoliubovtransformation}), 
impose conditions upon $A$ and $B$ matrices of the form
\begin{equation}
AA^T - BB^T = I\,\,\,\,\,\text{and}\,\,\,\,\,AB^T - BA^T = 0\,,
\label{condition}
\end{equation}
 where $I$ and $0$ are the identity and null $N \times N$ matrices, respectively. 
 A consequence of these equations is that $(A-B)(A^T+B^T)=I$, so $A\pm B$ is invertible
 with inverse $A^T\mp B^T$. 
Substituting Eq.~(\ref{Bogoliubovtransformation}) into Eq.~(\ref{eq:H-matrixform}), we obtain
\begin{equation}
\mathopen{:}H\mathclose{:} = \frac{1}{2}\begin{pmatrix} {\bf{b}}^{\dagger} & {\bf{b}}^T\end{pmatrix}
\left( \begin{array}{cc}
A^T & B^T \\
B^T &  A^T \end{array}\right) 
\left( \begin{array}{cc}
F & G \\
G &  F \end{array}\right)  
\left( \begin{array}{cc}
A & B \\
B &  A \end{array}\right)
\left( \begin{array}{c}
{\bf{b}} \\
{\bf{b}}^{\dagger T} \end{array}\right) -\frac{1}{2} \text{Tr}(F)\openone\,.
\label{Tbeforeimposediagnalizationcondition}
\end{equation}   
Now we impose a diagonalization condition 
 \begin{equation}
 \left( \begin{array}{cc}
A^T & B^T \\
B^T &  A^T \end{array}\right) 
\left( \begin{array}{cc}
F & G \\
G &  F \end{array}\right)  
\left( \begin{array}{cc}
A & B \\
B &  A \end{array}\right) = 
\left( \begin{array}{cc}
\Lambda & 0 \\
0 &  \Lambda \end{array}\right)\,,
\label{diagonalizationcondition}
 \end{equation}
 in Eq.~(\ref{Tbeforeimposediagnalizationcondition}), 
where $\Lambda = \text{diag}(\lambda_1,\ldots, \lambda_N)$. Using the canonical commutation relations for the $b_i$, 
we obtain
\begin{equation} 
\mathopen{:}H\mathclose{:}  = \sum_{i=1}^{N}\lambda_ib_i^{\dagger}b_i + C_{\text{shift}}\openone \,,
\label{Tafterimposediagnalizationcondition}
\end{equation} 
where 
\begin{equation}
\label{relationcshift}
C_{\text{shift}} = \frac{1}{2}\text{Tr}(\Lambda - F).
\end{equation}
It is clear that $\mathopen{:}H\mathclose{:}$ is diagonal in the orthonormal basis formed by vectors 
\begin{equation}
|{\bf n}\rangle_b= \left(\prod_{i=1}^{N} \frac{(b_i^\dagger)^{n_i}}{\sqrt{n_i!}}\right) |0\rangle_b \,
\end{equation} 
where ${\bf n}= (n_1,\ldots, n_N)$ with each $n_i$ a nonnegative occupation number, so that $b_i^{\dagger}b_i|{\bf n}\rangle_b = n_i|{\bf n}\rangle_b$ and $|0\rangle_b$ is annihilated by 
all the $b_i$. The eigenvalues are easily read off from  
\begin{equation}
\mathopen{:}H\mathclose{:} |{\bf n}\rangle_b = (n_i\lambda_i + C_{\text{shift}})|{\bf n}\rangle_b\,, 
\end{equation}
where the $i$-index runs from $1$ to $N$, and a sum on repeated indices is understood.
The operator $\mathopen{:}H\mathclose{:}$ is bounded from below provided that $\lambda_1,\ldots,\lambda_N$ are
all nonnegative, in which case $C_{\text{shift}}$ is the lowest eigenvalue. This gives a quantum inequality bound 
\begin{equation}
\langle \psi|\mathopen{:}H\mathclose{:}  |\psi\rangle \ge C_{\text{shift}}
\end{equation}
for all physical normalized states $\psi$. Note that $C_{\text{shift}}$ is both the lowest eigenvalue of the time-averaged stress
tensor operator, and the lower bound on its probability distribution, $P(x)$, so that $C_{\text{shift}} = -x_0$.

Let us return to the problem of achieving the diagonalization in practice. 
Noting that Eq.~(\ref{condition}) can be written in matrix notation as
\begin{equation}
 \left( \begin{array}{cc}
A & -B \\
-B &  A \end{array}\right) 
\left( \begin{array}{cc}
A^T & B^T \\
B^T &  A^T \end{array}\right)  
 = 
\left( \begin{array}{cc}
I & 0 \\
0 &  I \end{array}\right)\,,
\label{conditioninmatrixnotation}
 \end{equation}
we use the diagonalization condition, Eq.~(\ref{diagonalizationcondition}), to obtain
\begin{equation}
 \left( \begin{array}{cc}
F & G \\
G &  F \end{array}\right) 
\left( \begin{array}{cc}
A & B \\
B &  A \end{array}\right)  
 = 
 \left( \begin{array}{cc}
A & -B \\
-B &  A \end{array}\right) 
\left( \begin{array}{cc}
\Lambda & 0 \\
0 &  \Lambda \end{array}\right)
 =
\left( \begin{array}{cc}
A\Lambda & -B\Lambda \\
-B\Lambda &  A\Lambda \end{array}\right)\,,
\end{equation}
which is equivalent to a set of $2N$-equations to be solved for $A$, $B$, and $\Lambda$, given $F$ and $G$:
\begin{gather}
(F+G)(A+B)=(A-B)\Lambda\,,\label{FGplus1}\\
(F-G)(A-B)=(A+B)\Lambda\,.
\label{FGminus1}
\end{gather} 
A consequence of these equations and $(A\pm B)^{-1}=(A\mp B)^T$ is that
\begin{equation}
(A+B)^T (F+G) (A+B) = \Lambda = (A-B)^T (F-G) (A-B)
\end{equation}
and as we are interested in the case where $\Lambda$ is positive definite, it follows that 
a solution is only possible if both $F+G$ and $F-G$ are also positive definite. In this case, 
the equations can be solved as follows. First, because $F-G$ is positive, 
we may use  the Cholesky decomposition~\cite{Quarteroni2014} to find a real and invertible matrix $K$ such that $K^{\dagger}K = F-G$. 
The matrix $K(F+G)K^{\dagger}$ is real, symmetric and positive definite and therefore can be brought to diagonal form 
$U^{\dagger}K(F+G)K^{\dagger}U$ where all the diagonal entries are strictly positive and $U$ is a real orthogonal matrix. We then define
\begin{equation}
\Lambda=\sqrt{U^{\dagger}K(F+G)K^{\dagger}U}
\end{equation}

It may be verified (see Appendix A) that the solution to \eqref{FGplus1}
and \eqref{FGminus1} is given by $\Lambda$ together with
\begin{equation}
A = \frac{1}{2}(\Phi + \Psi)\,\,\,\,\,\,\text{and}\,\,\,\,\,B = \frac{1}{2}(\Phi - \Psi)\,,
\label{AandB}
\end{equation} 
where
\begin{equation}
\Phi = K^{\dagger}U\Lambda^{-1/2}\,\,\,\,\,\,\text{and}\,\,\,\,\,\Psi=(F+G)\Phi\Lambda^{-1}\,.
\end{equation} 

\subsection{Probabilities for particle sectors and outcomes for the single-mode case}
\label{onemodecase}
Now that we have the real matrices, $A$ and $B$, we want to express the original vacuum state, $|0\rangle_a$, as a linear combination of the eigenstates of $\overline{T}$, 
which are  linear combinations of the $|n_i\rangle_b$ in the new $b$-basis. First, we will develop the simplest case, a single mode, to obtain insight into the general case. 
The single mode case shows 
some interesting features which hold for the general case. In this case, $A$ and $B$ become $1 \times 1$ matrices, or real numbers. Express
\begin{equation}
|0\rangle_a = \sum_{n=0}^{\infty}C_n |n\rangle_b\\,
\label{linearcombination}
\end{equation}
where $C_n$ are coefficients to be determined. Apply the $a$-annihilation operator from the left and use the Bogoliubov transformation for the single mode case, 
Eq.~(\ref{Bogoliubovtransformation}), according to
\begin{align}
0 &= a|0\rangle_a  =\sum_{n=0}^{\infty}C_n (Ab + Bb^{\dagger})|n\rangle_b  \,,\\ 
& =   C_1A|0\rangle_b + \sum_{n=0}^{\infty}\left( C_{n+2}A\sqrt{n+2}|n+1\rangle_b + C_nB\sqrt{n+1}|n+1\rangle_b  \right)\,.
\label{Cn}
\end{align}
Now apply $(|0\rangle_b)^{\dagger}$ from the left to obtain $C_1 = 0$. Then, Eq.~(\ref{Cn}) becomes
\begin{gather}
\sum_{n=0}^{\infty}\left(C_{n+2}A\sqrt{n+2}+C_nB\sqrt{n+1} \right)|n+1\rangle_b  = 0\,.
\end{gather}
As the $|n\rangle_b$ form an orthonormal basis, we deduce  
\begin{equation}
C_{n+2}=-A^{-1}B\sqrt{\frac{n+1}{n+2}}C_n\,.
\label{Cnrecursive}
\end{equation} 
From this recursive expression and the fact that $C_1=0$, we have that all $C_n$ coefficients with odd-$n$ are zero. 
The $a$-vacuum is only connected with $|2n\rangle_b$ eigenstates of $\overline{T}$.
Let us make explicit this feature of the system and relabel $n$ by $2n$ in Eq.~(\ref{Cnrecursive}) and define
$\mathcal{M} \equiv A^{-1}B$  to obtain
\begin{equation}
C_{2n+2}=-\mathcal{M}\sqrt{\frac{2n+1}{2(n+1)}}C_{2n}\,.
\label{C2nrecursive}
\end{equation} 
It can be easily proved by induction that the general term in
Eq.~(\ref{C2nrecursive}) has the  form
\begin{equation}
C_{2n} 
=\left(-\frac{\mathcal{M}}{2}\right)^n\frac{\sqrt{(2n)!}}{n!}\,\mathcal{N}\,\,\,\,\,\text{for}\,\,\,\,\,n=0,1,2,3,\dots\,,
\label{C2nexpressionwithC0}
\end{equation}
where  $\mathcal{N} \equiv C_0$.
We apply the normalization condition to obtain $\mathcal{N}$ as
\begin{equation}
_{a}\hspace{-0.2 mm}\langle 0 | 0 \rangle_{a} = 1  = \sum_{n',n=0}^{\infty}  C_{2n'}^* C_{2n}\,\,  _{b}\hspace{-0.2 mm}\langle 2n' | 2n \rangle_{b}= 
\frac{|\mathcal{N}|^2}{\sqrt{1-\mathcal{M}^2}}\,,\,\,\,\,\,\text{with}\,\,\,\,\, |\mathcal{M}| < 1.
 \label{C0}
\end{equation} 
Then $|\mathcal{N}| = (1-\mathcal{M}^2)^{1/4}$. Here we have used Eq.~(\ref{linearcombination}) and the orthonormality property of the $a$-vacuum. 
Substituting the expression for $\mathcal{N}$ into Eq.~(\ref{C2nexpressionwithC0}), we have  
\begin{equation}
C_{2n}=\left(-\frac{\mathcal{M}}{2}\right)^n \frac{\sqrt{(2n)!}}{n!}(1-\mathcal{M}^2)^{1/4}\,,\,\,\,\,\,\text{for}\,\,\,\,\,n=0,1,2,3,\dots\,.
\end{equation}
As a result, the probability $P_{2n}$ of finding the original $a$-vacuum state in a specific $b$-state, $|2n\rangle_{b}$, and the corresponding eigenvalue of 
$\overline{T}$ , Eq.~(\ref{Tafterimposediagnalizationcondition}), are given by 
\begin{align}
P_{2n} & = \left\vert{b}\langle 2n | 0 \rangle_{a}\right\vert^2 = |C_{2n}|^2\,,\\
\overline{T} |2n\rangle_{b} & = \left( 2n\lambda + C_{\text{shift}}\right)|2n\rangle_{b}\,,
\end{align}
where $n = 0,1,2,3\dots$. From these equations, we see that the lowest possible outcome in a measurement of $\overline{T}$ is just $C_{\text{shift}}$, the $a$-vacuum state is 
only connected with  $2n$-particle sectors of the $b$-state, and the probability of finding the $a$-vacuum state in a specific $b$-state is concentrated in the lower
 particle number sectors. Indeed, the asymptotic expression for $P_{2n}$ decreases rapidly with $n$ according to 
\begin{equation}
P_{2n} \sim |\mathcal{N}|^2\frac{\mathcal{M}^{2n}}{\sqrt{\pi n}}\,,\,\,\,\,\,\text{for large } n \text{  and  } |\mathcal{M}| < 1.
\end{equation}
These three features of the single mode case hold for the general case that we now proceed to develop in the next subsection. 

\subsection{Probabilities for particle sectors and outcomes for the general case}
\label{multimodecase}
Express the $a$-vacuum state as a linear combination of $\psi_n$, where each $\psi_n$ belongs to the $n$-particle subspace for $b$-states, as follows
\begin{equation}
|0 \rangle_a = \sum_{n=0}^{\infty} \psi_{n}\,.
\end{equation}
Apply the $a_i$-annihilation operator from the left, use the Bogoliubov transformation, Eq.~(\ref{Bogoliubovtransformation}), and define again $\mathcal{M} \equiv A^{-1}B$ (now $\mathcal{M}$ 
is a matrix).  In detail,
\begin{equation}
b_k^\dagger \left[b_k + \mathcal{M}_{kj}b_j^{\dagger}\right]|0\rangle_a = b_k^\dagger (A^{-1})_{ki}(A_{ij}b_j+B_{ij}b_j^{\dagger}) |0\rangle_a b_k^\dagger=(A^{-1})_{ki} b_k^\dagger a_i|0\rangle_a = 0 \,,
\end{equation}
so  
\begin{align}
0  & = \sum_{n=0}^{\infty} \left[b_k^{\dagger}b_k + b_k^{\dagger}\mathcal{M}_{kj}b_j^{\dagger}\right]\psi_n\,,\\
& =   \psi_1 + \sum_{n=2}^{\infty}\left[ n\psi_n + ({\bf{b}}^{\dagger}\mathcal{M}{\bf{b}}^{\dagger T})\psi_{n-2}  \right]\,,
\label{Cnmultimode}
\end{align}
where we have used $b_k^{\dagger}b_k\psi_n = n\psi_n$ in the last line. 
The expression inside the bracket in Eq.~(\ref{Cnmultimode}) consists of $n$-particle terms with $n \geq 2$, thus we can only have a solution with $\psi_1 = 0$. That means that
\begin{equation}
\psi_n = -\frac{1}{n}\left( {\bf{b}}^{\dagger}\mathcal{M}{\bf{b}}^{\dagger T} \right)\psi_{n-2}\,\,\,\,\,\text{for n } \geq 2\,
\label{psi}
\end{equation}
and $\psi_1 = \psi_3 = \dots = \psi_{2n+1}=0$. We can rewrite Eq.~(\ref{psi}) by relabeling $n$ by $2n$ and expressing $\psi_{2n}$ in terms of $\psi_0$ as 
\begin{equation}
\psi_{2n} = \frac{(-1)^n}{2^n n!}\left( {\bf{b}}^{\dagger}\mathcal{M} {\bf{b}}^{\dagger T} \right)^n \psi_0\,.
\end{equation} 
Now, define $\psi_0 = \mathcal{N}|0\rangle_b$ to obtain
\begin{align}
|0\rangle_a & = \sum_{n=0}^{\infty} \psi_{2n} = \mathcal{N}\sum_{n=0}^{\infty} \left( -\frac{1}{2}  \right)^n \frac{1}{n!}\left( {\bf{b}}^{\dagger}M {\bf{b}}^{\dagger T}  \right)^n |0\rangle_b\,,\\
& = \mathcal{N}e^{-\frac{1}{2}{\bf{b}}^{\dagger}\mathcal{M} {\bf{b}}^{\dagger T}}|0\rangle_b\,,
\label{MBBequation}
\end{align}
where $\mathcal{N}$ is a normalization constant to be determined. Now, we diagonalize $\mathcal{M}$ such that $\mathcal{M} = S^T \Xi S$
with $S$ a real and orthogonal matrix and $\Xi = \text{diag}(\mu_i)$. Set  $c_i = S_{ij}b_j$ and 
$c_i^{\dagger}=S_{ij}b_j^{\dagger}$. They satisfy the bosonic commutation relations, because
\begin{align}
[c_i,c_k]&=[S_{ij}b_j,S_{kl}b_l]=S_{ij}S_{kl}[b_j,b_l]=0\,,\\
[c_i,c_k^{\dagger}]&=S_{ij}S_{kl}[b_j,b_l^{\dagger}]=(SS^T)_{ik}=\delta_{ik}\,.
\end{align}
Here we have used the commutation relation of $b$-operators and the orthogonality of $S$. Now note that we can rewrite the exponent in Eq.~(\ref{MBBequation}) using
\begin{equation}
{\bf{b}}^{\dagger}\mathcal{M}{\bf{b}}^{\dagger T}=b_i^{\dagger}\left( S^T \Xi S \right)_{ij} b_j^{\dagger} = \mu_l S_{li}S_{lj}b_i^{\dagger} b_j^{\dagger} =  \mu_lc_l^{\dagger}c_l^{\dagger}\,,
\end{equation}
where a sum on repeated indices is understood. Then the $a$-vacuum expressed in terms of the $b$-states, Eq.~(\ref{MBBequation}), becomes
\begin{equation}
|0\rangle_a = \mathcal{N} e^{-\frac{1}{2}\sum_{i}^{} \mu_i c_i^{\dagger} c_i^{\dagger}}|0\rangle_b\,. 
\end{equation}

The normalization constant $\mathcal{N}$ is calculated using the $a$-vacuum normalization and the definition for $c_i^{\dagger}$'s. For a single mode we have
\begin{align}
_{a}\langle 0 | 0 \rangle_{a} = 1 & = |\mathcal{N}|^2\left\| \sum_{n=0}^{\infty} \frac{-(\mu/2)^n}{n!}(c^{\dagger})^{2n} |0\rangle_b \right\|^2\,,\\
& = |\mathcal{N}|^2\left\|\sum_{n=0}^{\infty} \frac{(-\mu/2)^n}{n!}\sqrt{(2n)!} |2n \rangle_b\right\|^2\,,\\
& = |\mathcal{N}|^2 (1-\mu^2)^{-1/2}\,,\,\,\,\,\,\text{with}\,\,\,\,\,|\mu| < 1\,.
\end{align} 
Then $|\mathcal{N}| = (1-\mu^2)^{1/4}$. As  expected, we have recovered the result of the previous subsection, Eq.~(\ref{C0}), noting that for the single-mode case $\mu = \mathcal{M}$. 
For the multimode situation, we have
\begin{equation}
|\mathcal{N}| = \prod_i (1-\mu_i^2)^{1/4}\,.
\label{Nmultimode}
\end{equation}
The probability $P_{\lbrace n_i \rbrace}$ of finding the $a$-vacuum state in a specific $b$-state,  $|\left\lbrace n_k  \right\rbrace\rangle_b$, which now depends upon $N$-modes,
can be obtained from Taylor expanding the exponential in Eq.~(\ref{MBBequation}) according to
\begin{align}
P_{\lbrace n_k \rbrace} &= \left\vert_{b}\langle \left\lbrace n_k  \right\rbrace |0\rangle_a\right\vert^2  = 
\left\vert_{b}\langle \left\lbrace n_k  \right\rbrace | \mathcal{N}e^{-\frac{1}{2}{\bf{b}}^{\dagger}\mathcal{M}{\bf{b}}^{\dagger T}}|0\rangle_b\right\vert^2 \,,\\
& = \left\vert_{b}\langle \left\lbrace n_k  \right\rbrace | \mathcal{N}\left(  1 -\frac{1}{2}\sum_{i,j=1}^Nb_i^{\dagger}\mathcal{M}_{ij}b_j^{\dagger}+\dots\right)|0\rangle_b\right\vert^2 \,,\\
& = \left\vert_{b}\langle \left\lbrace n_k  \right\rbrace | \mathcal{N}\left(  |0\rangle_b -\frac{1}{\sqrt{2}}\sum_{i=1}^N \mathcal{M}_{ii}|2_i\rangle_b- 
\sum_{i<j}^N \mathcal{M}_{ij} |1_i 1_j\rangle_b + \dots \right)\right\vert^2\,.
\label{aux}
\end{align}
From this expression, we can determine, for example, the probability of finding the system in the $b$-vacuum state,  $P_{\lbrace{0\rbrace}}$, or in some configuration in the
 two-particle sector such as $P_{\lbrace{2_i\rbrace}}$ or $P_{\lbrace{1_i 1_j\rbrace}}$. These probabilities and the corresponding outcomes associated with a measurement 
 of $T$, Eq.~(\ref{Tafterimposediagnalizationcondition}), are specifically given by
\begin{align}
P_{\lbrace{0\rbrace}} & = |\mathcal{N}|^2\,,\,\,\,\,\,\text{and}\,\,\,\,\,\overline{T}|0\rangle_{b}= C_{\text{shift}}|0\rangle_{b}\,,\label{Pa}\\
P_{\lbrace{2_i\rbrace}} & = (1/2)|\mathcal{N}|^2|M_{ii}|^2\,,\,\,\,\,\,\text{and}\,\,\,\,\,\overline{T}|2_i\rangle_{b}= \left(2\lambda_i + C_{\text{shift}}\right)|2_i\rangle_{b}\,,\label{Pb}\\
P_{\lbrace{1_i 1_j\rbrace}} & =  |\mathcal{N}|^2|M_{ij}|^2\,,\,\,\,\,\,\text{and}\,\,\,\,\,\overline{T}|1_i1_j\rangle_{b}= \left(\lambda_i + \lambda_j + 
C_{\text{shift}}\right)|1_i1_j\rangle_{b}\,,\,\,\,\,\,\text{with}\,\,\,\,\,i < j\,.\label{Pc}
\end{align}
Here it is understood that $i,j$ run from $1$ to $N$. Now, we can re-express the normalization constant, $\mathcal{N}$, to obtain information about the total probability for 
each particle sector. We take Eq.~(\ref{Nmultimode}) and write the product as a determinant of the $\mathcal{M}$ matrix as
\begin{equation}
|\mathcal{N}|^2 = \prod_i(1-\mu_i^2)^{1/2}=\sqrt{\det(1-\mathcal{M}^2)}=e^{\frac{1}{2}\text{Tr}[\log(1-\mathcal{M}^2)]}\,,
\end{equation}
where we have used the well known formula $\det(W)=\exp\lbrace\text{Tr}[\log(W)]\rbrace$ for a given matrix, $W$. Expressing the $\log$-function as an infinite power series, 
and Taylor expanding the exponential, we can recognize the contribution for each particle sector as follows
\begin{align}
1 & = |\mathcal{N}|^2e^{-\frac{1}{2}\text{Tr}[\log(1-\mathcal{M}^2)]} = |\mathcal{N}|^2e^{\frac{1}{2}\sum_{n=1}^{\infty}\frac{\text{Tr}(\mathcal{M}^{2n})}{n}}\,,\\
& = |\mathcal{N}|^2  + |\mathcal{N}|^2\left[\frac{1}{2}\text{Tr}(\mathcal{M}^2)\right] + |\mathcal{N}|^2\left[ \frac{1}{4}\text{Tr}(\mathcal{M}^4)+\frac{1}{8}\text{Tr}^2(\mathcal{M}^2)\right]\,\nonumber \\
&\hspace{2 cm}+ 
|\mathcal{N}|^2\left[ \frac{1}{8}\text{Tr}(\mathcal{M}^2)\text{Tr}(\mathcal{M}^4)+\frac{1}{48}\text{Tr}^3(\mathcal{M}^2)+\frac{1}{6}\text{Tr}(\mathcal{M}^6) \right]+\mathcal{O}(\mathcal{M}^8)
\,.
\end{align}
Then, the contributions to the total probability of the $b$-vacuum and the two-particle sector, for instance, are 
$|\mathcal{N}|^2$ and $(1/2)|\mathcal{N}|^2\text{Tr}(\mathcal{M}^2)$, respectively. Each $2n$-particle sector contributes with terms having $2n$-factors of $\mathcal{M}$. 

\section{Massless scalar field in Minkowski spacetime}\label{sec:masslessfield}

\subsection{The square of the time derivative of the field}

We consider a minimally coupled massless scalar field, $\phi(x)$, in a four-dimensional Minkowski spacetime in spherical coordinates ($t,r,\theta,\varphi$), with the origin of the 
spherical polar coordinates placed at the fixed spatial point at which  $\mathopen{:}\dot{\phi}^2\mathclose{:}$ will be evaluated.  
The equation of motion is given by the usual wave equation
\begin{equation}
\partial_{\alpha} \partial^{\alpha} \phi(x) = 0\,.
\end{equation}
Solutions of this equation take the form~\cite{FordRoman1995}
\begin{equation}
f_{\omega lm}=\frac{g_{\omega l}(r)}{\sqrt{2\omega}}Y_{lm}(\theta,\varphi)e^{-i\omega t}\,,
\label{fwlmsolution}
\end{equation} 
where
\begin{equation}
g_{\omega l}(r)  = \omega \sqrt{\frac{2}{R}}j_l(\omega r)\,,
\end{equation}
and
\begin{equation}
1  =\int_{0}^{R}r^2 g_{\omega l}^2(r) dr\,.
\label{normalizedsphericalharmonic} 
\end{equation}
Here $j_l({\omega r})$ and $Y_{lm}(\theta,\varphi)$ are the spherical Bessel functions and the usual spherical harmonics, respectively. The normalization,
Eq.~(\ref{normalizedsphericalharmonic}), is carried out in a sphere of radius $R$. We set vanishing boundary conditions on the surface of the sphere by requiring
\begin{equation}
\phi(r)\vert_{r=R} = 0\,,
\end{equation}
which implies
\begin{equation}
\omega = \frac{z_{nl}}{R}\,,\,\,\,n = 1,2,\ldots\,. 
\label{omegabc}
\end{equation}
Here $z_{nl}$ is the $n$-th zero of the spherical Bessel function, $j_l$.

We expand the quantized field in terms of creation and annihilation operators, $a_{\omega lm}$ and $a_{\omega lm}^{\dagger}$, as
\begin{equation}
\phi(x) = \sum_{l = 0}^{\infty} \sum_{m = -l}^{l} \sum_{\omega}\left( a_{\omega lm}f_{\omega lm}+a_{\omega lm}^{\dagger}f_{\omega lm}^* \right)\,,
\label{expansionphi}
\end{equation} 
where a sum on $\omega$ is abbreviated notation for the sum on $n=1,2,\ldots$ with $\omega$ taking the values 
\eqref{omegabc} for the angular momentum sector $l$ in question. 

We want to calculate the time average of the normal-ordered quadratic operator $\mathopen{:}\dot{\phi}^2\mathclose{:}$ at fixed spatial point $r=0$ with sampling function $f(t)$, 
as in Eq.~(\ref{T}). Since all  $l \neq 0$ spherical Bessel functions vanish at $r=0$, we only have to consider the case $l=m=0$. Then, using 
$j_0(\omega r)=[\sin(\omega r)]/(\omega r)$ and $Y_{00}=1/\sqrt{4\pi}$ in Eq.~(\ref{fwlmsolution}), we have
\begin{equation}
f_{\omega 00}(t,r)=\frac{\sin(\omega r)}{r}\frac{e^{-i\omega t}}{\sqrt{4\pi \omega R}}\,,
\end{equation}
which, in the limit when $r \rightarrow 0$, becomes
\begin{equation}
f_{\omega 0 0}(t,0) = \sqrt{\frac{\omega}{4\pi R}}e^{-i\omega t}\,.
\label{fw00}
\end{equation}
Note that from the boundary conditions on the sphere, Eq.(~\ref{omegabc}), we have that $z_{n0}=n\pi$, so
\begin{equation}
\omega = \frac{n\pi}{R}\,,\,\,\,n = 1,2,\ldots\,. 
\label{bconditions}
\end{equation}
Making these simplifications in Eq.~(\ref{expansionphi}), taking the time derivative, and forming the Wick square, we obtain
\begin{equation}
\mathopen{:}\dot{\phi}^2\mathclose{:}(t,{\bf 0}) = 
\sum_{\omega}\sum_{\omega'}\frac{(\omega \omega')^{3/2}}{4\pi R}\left(  a_{\omega}^{\dagger}a_{\omega'}e^{i(\omega -\omega')t}-a_{\omega}a_{\omega'}e^{-i(\omega+\omega')t}+H.c.\right)\,,
\end{equation}
where $a_{\omega} \equiv a_{\omega 00}$, the sums run over the range given in  Eq.~\eqref{bconditions}, and $H.c.$ means hermitian conjugate. Convergence here should be understood 
in a distributional sense, so that when we now let $T= \mathopen{:}\dot{\phi}^2\mathclose{:}$ in Eq.~(\ref{T}), we find
\begin{equation}
\overline{T} =  \sum_{\omega}\sum_{\omega'} \frac{(\omega \omega')^{3/2}}{4\pi R}\left[ a_{\omega}^{\dagger}a_{\omega'}\hat{f}(\omega' - \omega)-a_{\omega}a_{\omega'}\hat{f}(\omega+\omega')
+H.c. \right]\,,
\end{equation} 
where $\hat{f}$ is the Fourier transform of the sampling function $f(t)$, Eq.~(\ref{FT}). 

\begin{figure}[ht]
\includegraphics[scale=0.45]{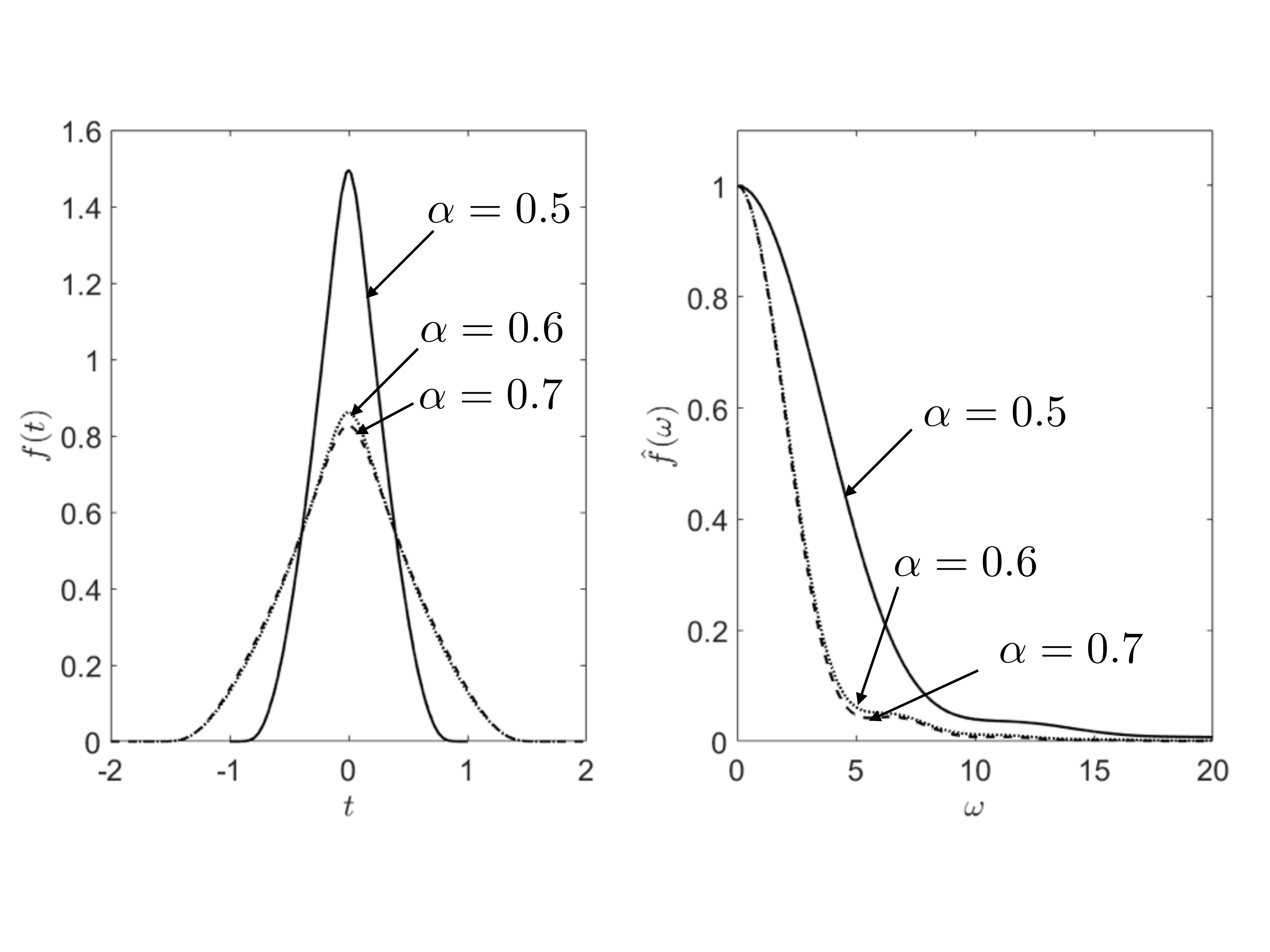}
\caption{Plots for the compactly supported function $f(t)$ (left) and its Fourier transform $\hat{f}(\omega)$ (right), for the cases of $\alpha=0.5$ (solid line),
$\alpha=0.6$ (dotted line), and $\alpha=0.7$ (dashed line). The values for $\delta$ used for each of these cases are, respectively, $0.5, 0.9,$ and $1.0$,
and units in which $\tau = 1$ are used.}
\label{ftandFT}
\end{figure}

We consider two different classes of sampling functions: the Lorentzian function whose Fourier transform is given by Eq.~(\ref{FTlorentzian}) ($\alpha =1$) and 
compactly supported functions whose Fourier transform has an asymptotic form when $\omega \tau \gg 1$ given by Eq.~(\ref{FTCSFunction}) ($\alpha\in\, (0,1)\,$). 
For this last case, we use a set of smooth, even, and nonnegative functions $f(t): \mathbb{R} \rightarrow [0,\infty)$ with compact support in $[-2\delta, 2\delta]$ and 
with Fourier transform given by (see Sect. IIA\&B of Ref.~\cite{Fewster:2015hga}
\begin{equation}
{\hat{f}(\omega) = \frac{\hat{H}^2\left(\omega\right)+\frac{1}{2}\left[ \hat{H}^2\left(\omega + \frac{\pi}{2\delta}\right) + 
\hat{H}^2\left(\omega - \frac{\pi}{2\delta}\right) \right]}{\hat{H}^2(0)+  \hat{H}^2\left(\frac{\pi}{2\delta}\right)}\,.}
\label{Fomega}
\end{equation}
Here $\hat{H}(\omega)$ is the Fourier transform of $H(t)= \varphi(t+\delta)\varphi(\delta-t)$, with $\varphi(t)$ being the inverse Laplace transform of 
$\tilde{\varphi}(p)=e^{-(p\tau)^{\alpha}}$. The Fourier transform $\hat{f}(\omega)$ is analytic, even, nonnegative and is normalized to one, $\hat{f}(0)=1$.  When $\omega \tau \gg 1$, 
$\hat{f}(\omega)$ has the asymptotic form given by Eq.~(\ref{FTCSFunction}) with
\begin{align}
& {\gamma = \frac{4\varphi^2(2\delta)}{\hat{H}^2(0)+\hat{H}^2(\frac{\pi}{2\delta})}\,,}\\
& {\beta = 2\cos\left(\frac{\pi \alpha}{2}\right)\,.}
\label{eq:beta-gamma}
\end{align}
Figure~\ref{ftandFT} plots the compactly supported function $f(t)$ and its Fourier transform $\hat{f}(\omega)$ for the cases of $\alpha=1/2$, 
$\alpha=0.6$, and $\alpha=0.7$. The plots for the $\alpha=1/2$ case agree with those in Figs.~4 and 5 in Ref.~\cite{Fewster:2015hga}, where the 
function and its Fourier transform were called $L(t)$ and  $\hat{L}(\omega)$, respectively.
It should be noted that $\tau$ is not the duration of the sampling period, which is $4\delta$, but rather sets the decay rate of the high frequency components in the sampling function and corresponds to a characteristic timescale of the switch-on and switch-off parts of $f(t)$. However it can serve as a proxy for the overall sampling time, within a set of functions related to $f$ by scaling. Using $\tau$ in this way also facilitates comparison with the Lorentzian function, for which the total sampling duration is infinite.
 
We define dimensionless variables $x_1 = \overline{T}(\tau^2)^2$ and $x_2 = \overline{T}(4\pi\tau^2)^2$ for the compactly supported functions and the Lorentzian function, respectively. 
(The difference in the numerical factors is to facilitate comparison with the results of Refs.~\cite{Fewster:2012ej} and \cite{Fewster:2015hga}, which used slightly different conventions.) 
Using the expression for $\omega$, Eq.~(\ref{bconditions}), these variables become
\begin{equation}
x_1  = \frac{1}{2}\sum_{r,s=1}^{\infty}\frac{\tau_0^4}{2\pi^2}(rs)^{3/2}\left[a_r^{\dagger}a_s \hat{f}(|r-s|\tau_0) -a_ra_s \hat{f}((r+s)\tau_0)+H.c.   \right]\,, \label{x1}
\end{equation}
and
\begin{equation}
x_2  = \frac{1}{2}\sum_{r,s=1}^{\infty}8\tau_0^4(rs)^{3/2}\left(a_r^{\dagger}a_se^{-|r-s|\tau_0}-a_ra_se^{-(r+s)\tau_0}+H.c.   \right)\,,\label{x2}
\end{equation}
 where we have defined
 \begin{equation}
   \tau_0 \equiv \pi \tau /R\,.
   \label{eq:tau0}
 \end{equation}  
   Note that the  expressions for $x_1$ and $x_2$ have the form of Eq.~(\ref{Trewritten}).
Thus, the matrices $F$ and $G$ for the case of a compactly supported function are
\begin{equation}
{F_{rs} =\frac{\tau_0^4}{2\pi^2}(rs)^{3/2}\hat{f}(|r-s|\tau_0) \,\,\,\,\,\text{and}\,\,\,\,\,G_{rs} = -\frac{\tau_0^4}{2\pi^2}(rs)^{3/2}\hat{f}((r+s)\tau_0)\,.}
\end{equation} 
Similarly, those for the case of a Lorentzian sampling function are
\begin{equation}
F_{rs} = 8\tau_0^4(rs)^{3/2}e^{-|r-s|\tau_0}\,\,\,\,\,\text{and}\,\,\,\,\, G_{rs} = -8\tau_0^4(rs)^{3/2}e^{-(r+s)\tau_0}\,.
\end{equation}
The $F$ and $G$ matrices are all that we need to calculate, for a given number of modes, the probability distribution and the cumulative probability distribution associated with a measurement 
of $x_1$ or $x_2$. 

\subsection{Numerical results for the cumulative probability distribution function and tail for large fluctuations} 
\label{CDF}

Here we explain the general features of the numerical calculation that we carry out to calculate the probability, $P(x)$, and cumulative probability distribution function, $P_>(x)$, for the two cases 
mentioned above. Here $x$ 
denotes either $x_1$ or $x_2$, defined in Eqs.~(\ref{x1}) and ~(\ref{x2}). For a given number of modes, we calculate all possible outcomes in a measurement of $x$ up to and including the 6-particle 
sector, except for the following outcomes which have been omitted:
\begin{align}
& \lambda_i + \lambda_j + \lambda_k + \lambda_l + \lambda_m + \lambda_n + C_{\text{shift}}\,,\label{TakeoutOutcome1}\\
& 2\lambda_i + \lambda_j + \lambda_k + \lambda_l + \lambda_m  + C_{\text{shift}}\,.\label{TakeoutOutcome2}
\end{align}  
Recall that the $\lambda_i $ are the one-particle eigenvalues which appear in Eq.~\ref{Tafterimposediagnalizationcondition}.   
Here it is understood all indices are different in these expressions. These outcomes were not included due to
the large number of operations that they would entail. For example, the outcome with six different eigenvalues, Eq.~(\ref{TakeoutOutcome1}), would involve about $10^9$ operations 
for the case of 100 modes. All probabilities and outcomes included in the calculation are listed explicitly in Appendix B. We build the cumulative distribution  $P_{>}(x)$ by  adding 
the probabilities of outcomes, $P_{\lbrace n_i \rbrace}$, from Eq.~(\ref{aux}),  which are sorted from the lowest to the largest value of $x$. 

The number of modes and the value 
for $\tau_0$ are crucial in determining the quality of the $P_>(x)$-curve. Recall that we have standing waves, Eq.~(\ref{fwlmsolution}), inside a sphere of radius $R$, which
is related to $\tau$ and $\tau_0$ by Eq.~(\ref{eq:tau0}), and that the sampling timescale $\tau$ is defined for the Lorentzian function in Eq.~(\ref{FTlorentzian}), and  for the 
compactly supported functions in Eqs.~(\ref{FTCSFunction}) and (\ref{eq:beta-gamma}). For a fixed  characteristic timescale, $\tau$, the radius of the sphere is inversely proportional to 
dimensionless variable $\tau_0$. For a given number of modes, if the size of the sphere is too large, there will not be enough data in the tail ($x \gg 1$) of the 
$P_>(x)$-curve to perform a reliable fit. By contrast, if the size of the sphere is too small, 
the  $P_>(x)$-curve will not be smooth, showing a step-like behavior. For the  compactly supported functions, we also have to 
determine  values for $\delta$, which defines the support of the sampling function $f(t)$, i.e., the duration of the sampling. We choose these values to be slightly larger than the first maximum 
of the corresponding $\varphi(t)$, and the results are given in Table~\ref{TableParticleSectors}, working in units where $\tau=1$. Then the sphere radius
 $R= \tau_0/\pi$ gives values $1.14$, $0.64$ and $0.64$ for $\alpha = 0.5, 0.6, 0.7$, respectively, for the values of $\tau_0$ considered. 
Note that for $\alpha = 0.5$ we have $R>2\delta$, which means that the total sampling time
is less than the time taken for light to travel to the boundary and back. Accordingly,  
the numerics ought to give a good approximation to sampling in Minkowski space; this is an instance of local covariance, which has a number of applications to quantum inequalities~\cite{Fewster:2006kt}.
By contrast, in the other two remaining cases we have $R<2\delta$, so the sampling process can be sensitive to the presence of the bounding sphere.
The reduced values of $\delta$ used for $\alpha=0.6,0.7$ were required to obtain numerical stability.

\begin{table}[ht]
\begin{threeparttable}
\caption{Numerical results for the parameters of the $P_>(x)$-curves illustrated in  Fig.~\ref{CDFcurves}, for the case of compactly supported functions with 
different values of $\alpha$ and for the Lorentzian function. Units in which $\tau=1$ have been adopted.
Here values of $P_>(x)$ for different particle sectors are calculated adding all probabilities for all possible outcomes for the given sector as is indicated in 
Table~\ref{TablePxOutcomes}. Since $x_{max}$ is the maximum value obtained in a measurement of $x$ for a given number of modes and size of the sphere, the expression 
$[1- P_> (x_{max})]$ gives us the loss of probability.} 
\label{TableParticleSectors} 
\centering 
\begin{tabular}{c |   c |   c |   c |   c} 
\hline\hline 
 & \boldmath{\bf{ $P_> (x_1)$}} &\boldmath{\bf{ $P_> (x_1)$}} & \boldmath{\bf{  $P_> (x_1)$}} &\boldmath{\bf{  $P_> (x_2)$}}  \\ [0.2ex] 
 & \boldmath${\bf{\alpha = 0.5}}$ &\boldmath${\bf{\alpha = 0.6}}$ &  \boldmath${\bf{\alpha = 0.7}}$ & \bf{Lorentzian} \\ [0.2ex]
\hline 
$\bf{Modes}$ & $120$ & $120$  & $120$ & $140$\\
\hline
$\bf{Points}$ & $\mathcal{O}(10^9)$ & $\mathcal{O}(10^9)$ & $\mathcal{O}(10^9)$ & $\mathcal{O}(10^9)$\\
\hline
\boldmath$\bf{\delta}$ & $0.5$ & $0.9$ & $1.0$ & $\--$\\
\hline
\boldmath$\bf{\gamma}$ & $2.9324$ & $1.0433$ & $0.5235$ & $1$\\
\hline
\boldmath$\bf{\beta}$ & $1.4142$ & $1.1756$ & $0.9080$ & $1$\\
\hline
\boldmath$\bf{\textit{f}~(0)}$ & $1.4990$ & $0.8616$ & $0.8274$ & $0.6366$\\
\hline
\boldmath$\bf{\tau_0}$ & $3.5725$ & $2.0$ & $2.0$ & $0.2$\\
\hline
\boldmath$x_{max}$ & $\mathcal{O}( 10^{8})$ & $\mathcal{O}(10^7)$ &  $\mathcal{O}( 10^7)$ & $\mathcal{O}(10^6)$ \\
\hline
$\bf{C_{\text{shift}}}$ & $-7.81613 \cdot 10^{-2}$ & $-1.48420\cdot 10^{-2}$ &  $-1.37113\cdot 10^{-2}$ & $-5.93338 \cdot 10^{-2}$ \\
\hline
$\bf{Vacuum}$  & $9.88503\cdot 10^{-1}$ & $9.72841\cdot 10^{-1}$ & $9.71898\cdot 10^{-1}$& $9.70277\cdot 10^{-1}$ \\ 
\hline 
$\bf{2}^{\bf{nd}} \bf{sector}$  & $1.13068 \cdot 10^{-2}$& $2.61008 \cdot 10^{-2}$ & $2.69537 \cdot 10^{-2}$ & $2.87007 \cdot 10^{-2}$\\
\hline 
$\bf{4}^{\bf{nd}} \bf{sector}$ & $1.86704 \cdot 10^{-4}$& $1.01218 \cdot 10^{-3}$ & $1.09604 \cdot 10^{-3}$& $9.48946 \cdot 10^{-4}$  \\
\hline 
$\bf{6}^{\bf{nd}} \bf{sector}$  & $3.44828 \cdot 10^{-6}$& $4.38949 \cdot 10^{-5}$ & $4.97286 \cdot 10^{-5}$& $2.97518 \cdot 10^{-5}$ \\
\hline 
\boldmath$\left[1- P_> (x_{max})\right]$ & $6.83316 \cdot 10^{-8}$ & $2.09890 \cdot 10^{-6}$ & $2.49384 \cdot 10^{-6}$ & $4.37397 \cdot 10^{-5}$\\
\hline\hline 
\end{tabular}
    \end{threeparttable}
\end{table}

\begin{figure}
\centering
\begin{minipage}{0.5\textwidth}
  \centering
  \includegraphics[width=1\linewidth]{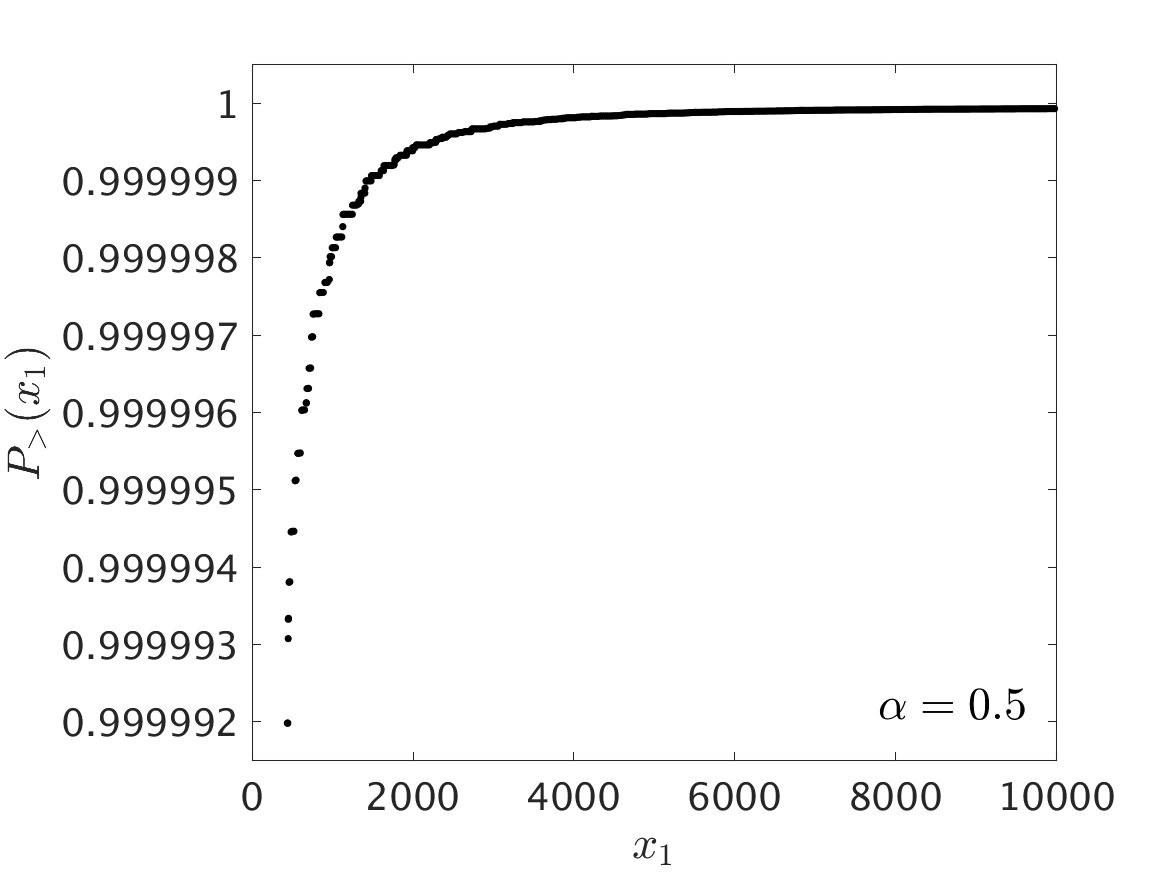}
  \label{fig:test1}
\end{minipage}%
\begin{minipage}{0.5\textwidth}
  \centering
  \includegraphics[width=1\linewidth]{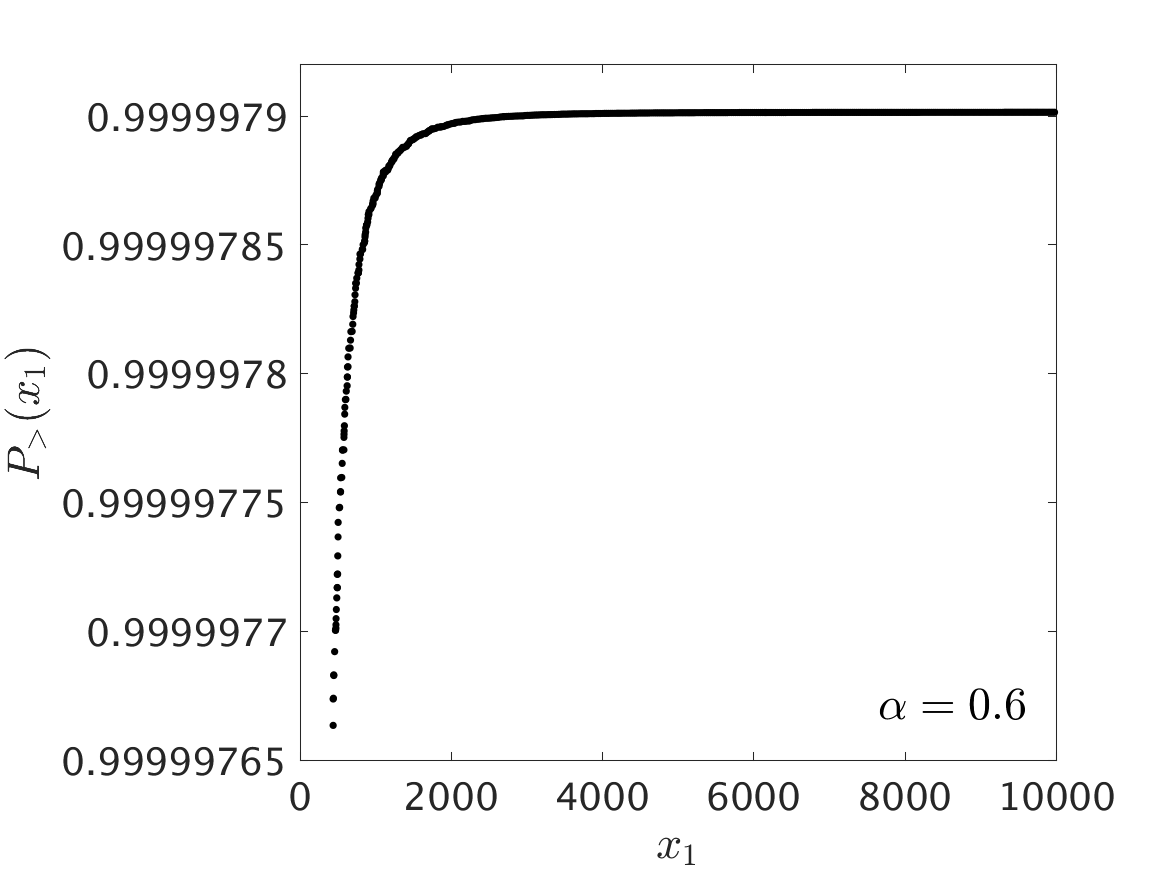}
  \label{fig:test2}
\end{minipage}
\centering
\begin{minipage}{0.5\textwidth}
  \centering
  \includegraphics[width=1\linewidth]{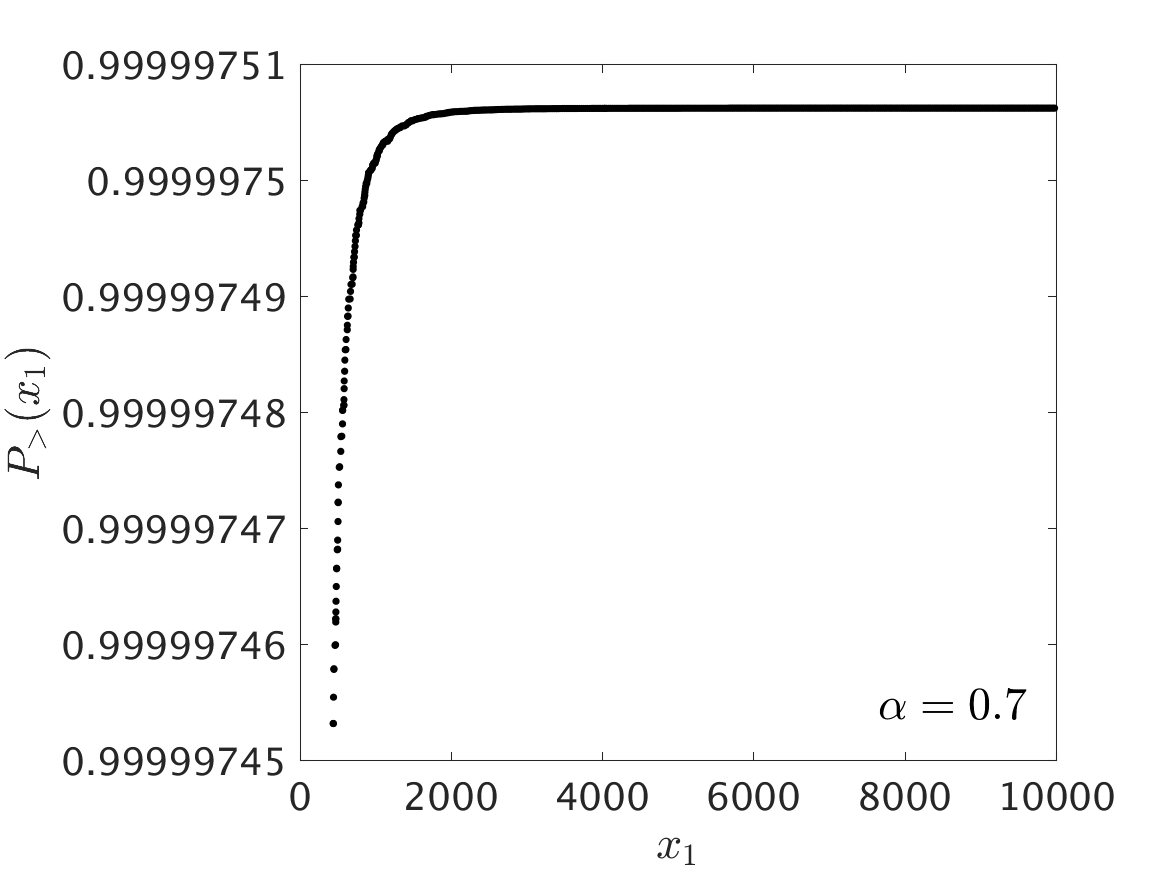}
  \label{fig:test3}
\end{minipage}%
\begin{minipage}{0.5\textwidth}
  \centering
  \includegraphics[width=1\linewidth]{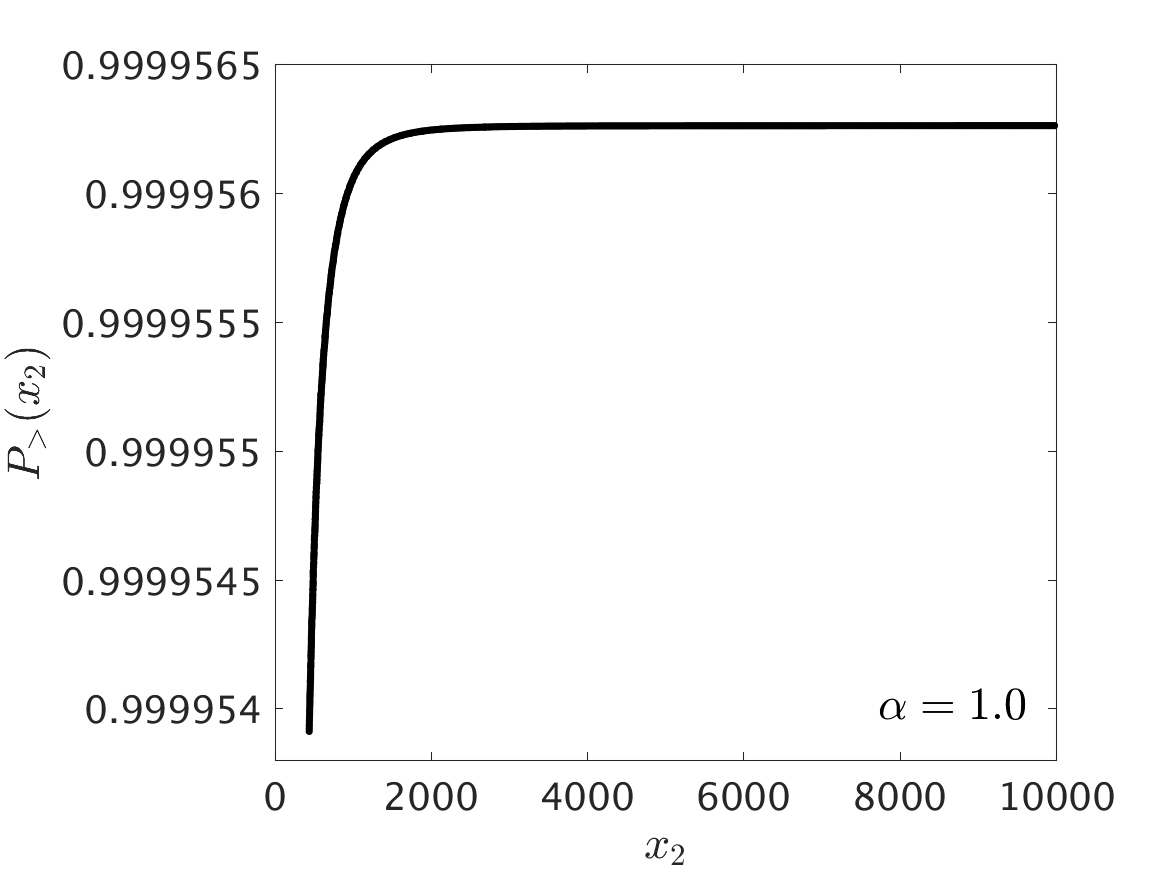}
  \label{fig:test4}
\end{minipage}
\caption{$P_>(x)$-curves for the case of compactly supported functions with decay parameters ($\alpha = 0.5, 0.6, 0.7$) and the Lorentzian function ($\alpha = 1$) 
for the range $450 \lesssim x \lesssim 10000$.  Additional information is shown in detail in Table~\ref{TableParticleSectors}.}
\label{CDFcurves}
\end{figure}

We build $P_>(x)$-curves for compactly supported functions whose Fourier transform is given by Eq.~(\ref{Fomega}) with three different values of the decay parameter, 
$\alpha = \left( 0.5, 0.6,  0.7 \right)$, and the Lorentzian function. Table~\ref{TableParticleSectors} summarizes the main characteristic of these curves which are shown 
by Fig.~\ref{CDFcurves} for the range $450 \lesssim x \lesssim 10000$.  All curves are smooth, show the presence of large vacuum fluctuations 
($x \gg 1$), and have sufficient amount of data to carry out the subsequent fit procedure. Recall that the original $a$-vacuum state is expressed in terms of a linear 
combination of $b$-states which are eigenstates of $x$. As expected, the most likely $b$-state is the $b$-vacuum state  and the $P_>(x)$-curves are bounded  below 
by the value $C_{\text{shift}} = -x_0 < 0$.  The loss of probability for each case is given by $[1-P_>(x_{max})]$, where $x_{max}$ is the maximum value obtained in a measurement of $x$ for a given number of 
modes and size of the sphere.  All analyzed cases show a small loss of probability of the order of $10^{-5}$ or less.  This small loss of probability  indicates that the
outcomes which have been included provide a reasonable approximation for   $P_>(x)$.   

Our calculated values of the lower bound $C_{\text{shift}}$ can be compared with results from other approaches. In the case of the Lorentzian function, 
our calculated value $C_{\text{shift}}=-0.0593338$, is of the order of the predicted value from the analysis using high moments, $x_0 = -0.0236$~\cite{Fewster:2012ej} 
and well within the (non-optimal) theoretical bound $C_{\text{shift}}\ge -27/128=-0.211$ given by the method of~\cite{Fewster:1998pu}. For a general compactly supported 
test function $f$, the theoretical bound is
\begin{equation}\label{eq:Cshift_theoretical}
C_{\text{shift}}\ge - \frac{\tau^4}{16\pi^2}\int_{-\infty}^\infty (f^{1/2\prime\prime}(t))^2\, dt 
\end{equation}
which can be obtained by setting $p=\sqrt{\omega}$ in Eq.~(3.11) of \cite{Fewster:1998pu}.
For the case $\alpha=0.5$, the integral on the right-hand side of~\eqref{eq:Cshift_theoretical} can be evaluated numerically and yields the bound $C_{\text{shift}}\ge -0.3592$. 
Our calculated value  $C_{\text{shift}}=-0.0781613$ is therefore consistent with the theoretical bound and indicates that the latter bound is weaker than the sharpest possible 
bound by a factor of approximately $4.6$. This result is broadly in line with Dawson's computations~\cite{Dawson2006}, where a ratio of about $3$ was found. Note that Dawson used a toroidal spatial geometry rather than a ball and a squared Lorentzian sampling function of infinite duration, so one would not expect an exact match with our results.

Since we want to test the predicted behavior of the cumulative probability distribution for large fluctuations in vacuum, we focus on the tail of each $P_>(x)$-curve and 
propose a trial function inspired by Eq.~(\ref{asymptoticPX}). Specifically,
\begin{equation}
P_>(x;\hat{\theta}) =  p_1 - \frac{c_0a^{-(1+b)/c}}{c}\Gamma\left( \frac{1+b}{c},ax^c \right)\,.
\label{bestfit}
\end{equation} 
Here $\hat{\theta} = (p_1, a, b, c, c_0)$ are the five free parameters to be determined through the usual process of best-fitting.
We fit the numerical data to this trial function. Producing a $P_>(x)$-curve implies propagating errors from the successive sum of the $P_{\lbrace n_i \rbrace}$, defined 
in Eq.~(\ref{aux}), but errors coming from the diagonalization procedure are mostly dominated by the error in $|N|^2$, from the vacuum sector. 
Constructing the tail of each $P_>(x)$-curve entails dealing with $10^6$ data points. To make the fitting-procedure possible in a reasonable  time, we bin the data as follows. 
Let $N$ be the total number of data points.  We split this set in several subsets $N_i$, where $N =\sum_{i=1}^j N_i$ and $j$ is the total number of subsets.  Consider one  
subset of values of $x$ and the associated values of $P_>(x)$, $N_i = \lbrace (x_1,P_>(x_1)), (x_2,P_>(x_2)), \dots , (x_{N_i},P_>(x_{N_i})) \rbrace$. Next replace it by 
the averaged values $\overline{N}_i =(\bar{x}_i,\overline{P}_>(\bar{x}_i))$, where $\bar{x}_i = \sum_{k=1}^{N_i} x_k/N_i$ and 
$\overline{P}_>(\bar{x}_i) = \sum_{k=1}^{N_i} P_>(x_k) / N_i$. The size of the subset is taken to depend on the steepness of the  $P_>(x)$-curve. The steeper this curve, 
the smaller is $N_i$. This procedure ensures that the best fit to the set of averaged values represents  a good fit of the original curve. 
The $10^6$ data points are typically divided into about $10^3$ bins. The values of $N_i$, the number of points per bin, range from about $10^2$ at the smaller values of
$x$ to about $10^4$ at the larger values.

The fitting procedure is 
based on the least-squares method to find the specific set of values of parameters which minimize the error variance. We name this specific set as 
$\theta^* = (p_1^*, a^*, b^*, c^*, c_0^*)$. The estimation of the error variance, $s^2$, is given by   
\begin{equation}
s^2 = \frac{1}{(j-5)}\sum_{i=1}^j {\frac{\left[\overline{P}_>(\bar{x}_i)-P_>(\bar{x}_i;\hat{\theta})\right]^2}{(N/N_i)}} \,,
\label{errorvariance}
\end{equation} 
 where $(j-5)$ is the number of degrees of freedom, $\overline{P}_>(\bar{x}_i)$ is the $i$th value of the 
 averaged $\overline{P}_>(\bar{x})$-curve, $P_>(\bar{x}_i;\hat{\theta})$ is the $i$th value of the
 fitting-curve. Note that we are weighting each $i$th value of the square of the residuals, $[~\overline{P}_>(\bar{x})-P_>(\bar{x}_i;\hat{\theta})~]^2$,
by the ratio  $(N/N_i)$. This gives a greater weight to the larger subsets. We have also assumed that the error in the values associated with the $j$ different subsets is the same.
This allows us to directly sum the squares of the residuals over the various subsets. If the errors of the different subsets are different, then weight factors for each subset would 
be needed.

\begin{figure}[ht]
\centering
\includegraphics[scale=0.65]{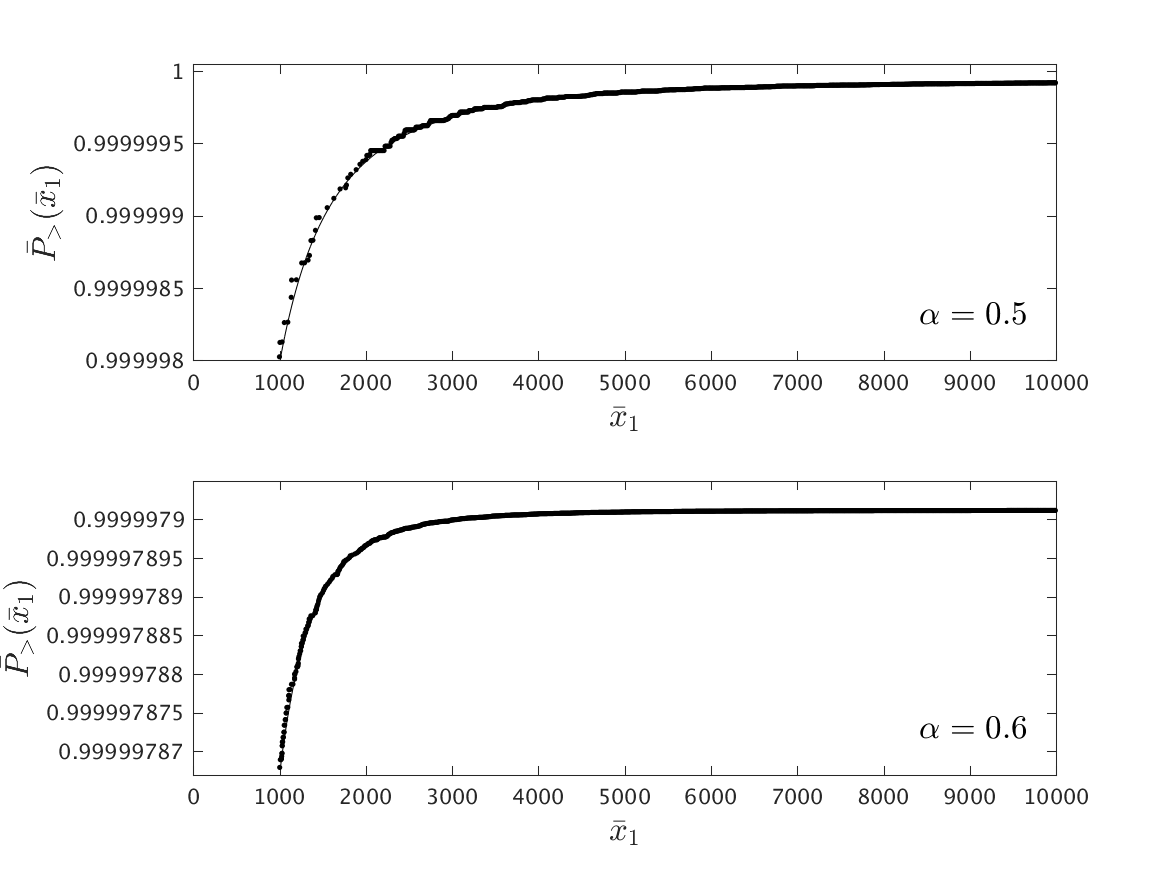}
\includegraphics[scale=0.65]{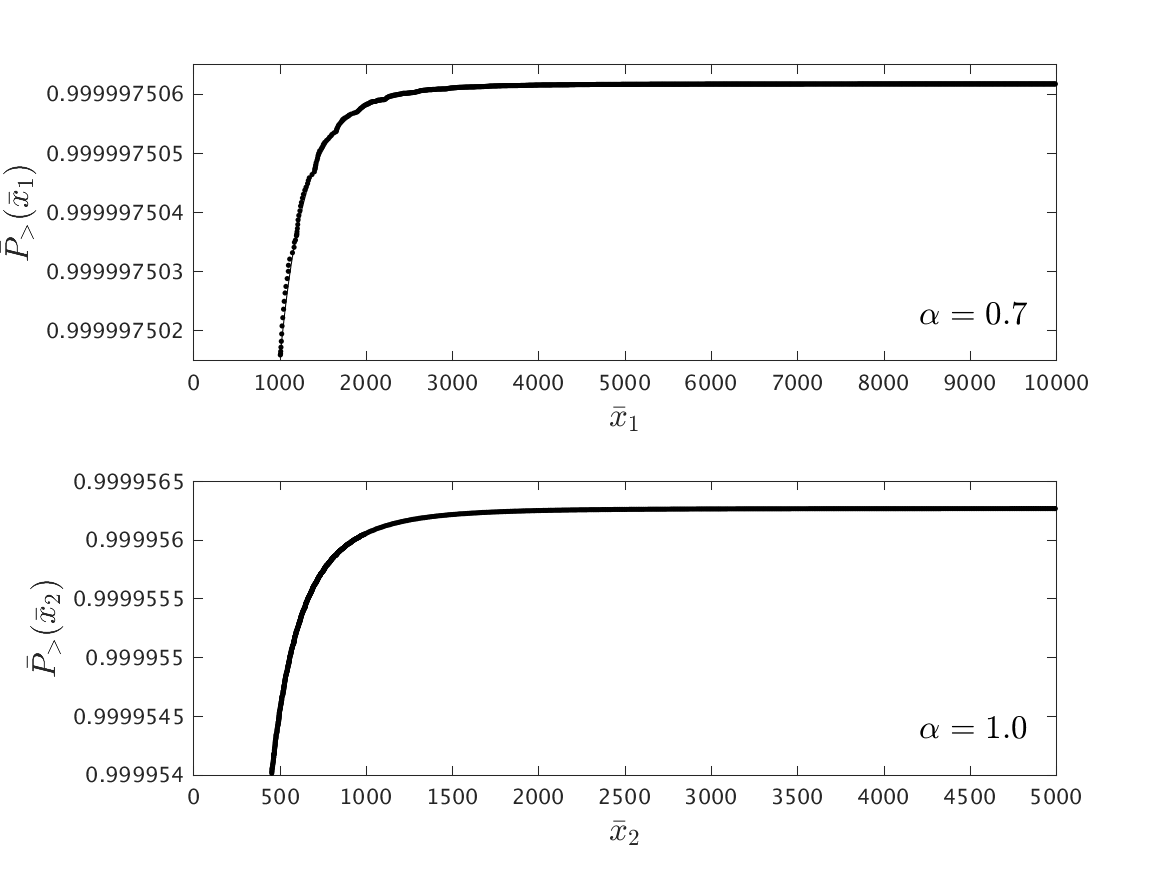}
\caption{Best fitting using Eq.~(\ref{bestfit}) to reproduce the $\overline{P}_>(\bar{x})$-curve for the cases of compactly supported functions with different decay parameters  
(range of $1000 \lesssim x \lesssim 10000$) and for the Lorentzian function (range of $400 \lesssim x \lesssim 5000$). In each case, the dots and the line correspond to the 
$\overline{P}_>(\bar{x})$-curve and its fit, respectively. For the case of the Lorentzian function, dots and line are indistinguishable on the scale shown. }
\label{Plotfitting}
\end{figure}

Table~\ref{parametersalphas} summarizes the statistical information obtained by the  best-fitting procedure for each case which includes the estimate value for parameters 
and their respective standard errors (only from statistical sources). Figure~\ref{Plotfitting} shows the $\overline{P}_>(\bar{x})$-curves with their respective
best fits to the trial function, Eq.~(\ref{bestfit}). In the case of the Lorentzian function, the  $\overline{P}_>(\bar{x})$-curve and its respective fit are indistinguishable on the 
scale shown. Figure~\ref{Plotfitting} shows that the diagonalization procedure is able to reproduce smooth tails for all the cases considered, which are well fitted by the trial 
function given by an incomplete gamma function, Eq.~(\ref{bestfit}).
The  variance of the fits are small in comparison to the variation 
of the $\overline{P}_>(\bar{x})$-curves. For example, for the case of the compactly supported function with $\alpha = 0.7$, we have $s^2 \sim \mathcal{O}(10^{-22})$ 
but the change of the $\overline{P}_>(\bar{x})$-curve over the range plotted in Fig.~\ref{Plotfitting} is the order of $10^{-9}$.

\begin{table}[ht]
\begin{center}
\caption{Parameters obtained from the best fit of Eq.~(\ref{bestfit}) for compactly supported functions with different values of $\alpha$, and for the Lorentzian function.}
\label{parametersalphas}
\begin{tabular}{c|c|c|c|c|c|c}
\multicolumn{1}{c|}{} & \multicolumn{3}{|c|}{\boldmath$\alpha = 0.5 \,\,(s^2 \sim 10^{-18})$} & \multicolumn{3}{|c}{\boldmath$\alpha = 0.6 \,\,(s^2 \sim 10^{-21})$} \\
\hline\hline
& Estimate & Standard Error & Theoretical~\cite{Fewster:2015hga}  & Estimate& Standard Error & Theoretical~\cite{Fewster:2015hga}  \\
\hline
\boldmath$p_1^*$ & $1$ & $9.86890 \cdot 10^{-10}$ & $1$ & $1$ & $3.82057 \cdot 10^{-12}$ & $1$ \\
\hline
\boldmath$a^*$ & $3.21574$ & $0.26916$ & $3.19965$ & $2.86707$ & $3.09190 \cdot 10^{-3}$ & $3.04545$ \\
\hline
\boldmath$b^*$ & $-0.64913$ & $6.74595 \cdot 10^{-2}$ & $-1$ & $-1.29164$ & $1.94800 \cdot 10^{-3}$ & $-1.13333$ \\
\hline
\boldmath$c^*$ & $0.17368$ & $6.21754 \cdot 10^{-3}$ & $0.16667$ & $0.198625$ & $1.74722 \cdot 10^{-4}$ & $0.2$ \\
\hline
\boldmath$c_0^*$ & $1.24953 \cdot 10^{-2}$ & $6.17359 \cdot 10^{-3}$ & $4.84678 \cdot 10^{-2}$ & $5.52294 \cdot 10^{-2}$ & $8.36918 \cdot 10^{-4}$ & $1.57857 \cdot 10^{-2}$\\
\hline
\multicolumn{1}{c|}{} & \multicolumn{3}{|c|}{\boldmath$\alpha = 0.7 \,\,(s^2 \sim 10^{-22})$} & \multicolumn{3}{|c}{\boldmath${\bf{Lorentzian}} \,\,(s^2 \sim 10^{-17})$} \\
\hline\hline
& Estimate & Standard Error & Theoretical~\cite{Fewster:2015hga} & Estimate & Standard Error & Theoretical~\cite{Fewster:2012ej}  \\
\hline
\boldmath$p_1^*$ & $1$ & $6.87424 \cdot 10^{-13}$ & $1$ & $0.99996$ & $5.44678 \cdot 10^{-11}$ & $1$\\
\hline
\boldmath$a^*$ & $2.74969$ & $1.18528 \cdot 10^{-3}$ & $2.47920$ & $1.04998$ & $1.19509 \cdot 10^{-2}$ & $0.667749$\\
\hline
\boldmath$b^*$ & $-1.17210$ & $5.71700 \cdot 10^{-4}$ & $-1.26667$ & $-1.14578$ & $9.33892 \cdot 10^{-3}$ & $-2$\\
\hline
\boldmath$c^*$ & $0.228107$ & $1.15423 \cdot 10^{-4}$ & $0.23333$ & $0.315336$ & $1.07643 \cdot 10^{-3}$ & $0.333333$\\
\hline
\boldmath$c_0^*$ & $3.05954 \cdot 10^{-2}$ & $3.22838 \cdot 10^{-4}$ & $5.44308 \cdot 10^{-3}$ & $2.08459 \cdot 10^{-2}$ & $8.84006 \cdot 10^{-4}$ & $0.477696$\\
\hline\hline
\end{tabular}
\end{center}
\end{table}

All values of parameters obtained through the best-fitting procedure agree reasonably well with the predicted ones from the high moments 
approach~\cite{Fewster:2012ej, Fewster:2015hga}, except for those for the $c_0$ parameter. The deviations of the fitted values for this parameter from
to the predicted values, which are of $~\mathcal{O}(1)$ or less, are probably caused by the use a finite number of  modes and finite size of the sphere. 
The most important parameter to be evaluated is $c$, because it is related to the rate of decrease of the probability distribution for large fluctuations, Eq.~(\ref{asymptotic}). Recall that $c = \alpha/3$, 
where $\alpha \in\, (0,1)$ is the decay parameter for the family of compactly supported functions with Fourier transform given by Eq.~(\ref{Fomega}). For the case of a Lorentzian function we have 
$c=1/3$. The values of the $c$ parameter obtained for each case agree very well with the predicted ones, with a percentage error less than $6\%$. For instance, for the case of a compactly supported 
function with $\alpha = 0.6$, the percentage error is about $0.69\%$.  In complete agreement with previous results based on the high moments analysis~\cite{Fewster:2012ej, Fewster:2015hga}, our 
results confirm the fact that averaging over a finite time interval compactly supported functions results in a  probability distribution which falls more slowly than for the case of the Lorentzian function, 
and both fall more slowly than exponentially.

\section{Summary and Discussion}
\label{sec:summary}

Large vacuum fluctuations of quantum stress tensor operators can have a variety of physical effects such as production of gravity waves in inflationary models~\cite{Wu:2011gk}, 
fluctuations of the light propagation speed in nonlinear materials~\cite{Bessa:2014pya, Bessa:2016uqb}, and enhancing barrier penetration of charged or polarizable 
particles~\cite{Huang:2015lea, Huang:2016kmx}. These quantum fluctuations can be studied through the analysis of the probability distribution for the time or spacetime averaged 
operator in Minkowski spacetime. The asymptotic behavior 
of the probability distribution can be inferred by studying the moments of the normal ordered operator. The study of several normal-ordered quadratic operators time averaged with a Lorentzian 
function~\cite{Fewster:2012ej} or compactly supported functions~\cite{Fewster:2015hga} predict an asymptotic form of the probability distribution for large vacuum fluctuations $x$ given by 
$P(x) \sim c_0 x^b e^{-ax^c}$, Eq.~(\ref{asymptotic}), where $x$ is a dimensionless measure of the quadratic operator. This form leads to an asymptotic form for the cumulative probability distribution 
given by an incomplete gamma function, Eq.~(\ref{asymptoticPX}). Here $c_0, a, b,$ and $c$ are constants which depend on the sampling function used to take the time average. 
The $c$-parameter is the most important one, and defines the rate of decrease of the tail of the probability distribution. For the family of compactly supported functions with  
asymptotic Fourier transforms given by Eq.~(\ref{FTCSFunction}), where  $0 < \alpha <1$, we have $c = \alpha/3$. For the case of a Lorentzian function,
Eq.~(\ref{FTlorentzian}), we have $\alpha=1$ and $c=1/3$. The smaller $\alpha$, the smaller the rate of decrease of the tail and greater the probability of large fluctuations. The value of $\alpha$ is 
related to the rate of switch-on and switch-off of compactly supported functions. 
 
 In the present paper, we have developed a method which is independent of the moments approach for the study of the probability distribution for quantum vacuum fluctuations of a time averaged 
 quantum stress tensor operator, $\overline{T}$, in Eq.~(\ref{T}). Since the vacuum state is not usually an eigenstate of $\overline{T}$, we  diagonalize this operator through a change of basis. 
 Expressing the vacuum state in terms of the new basis in which $\overline{T}$ is diagonal, we are able to calculate the probability distribution, $P(x)$ and the cumulative probability  
 distribution function, $P_>(x)$ for obtaining a specific result in a measurement of $\overline{T}$. Specifically, we work with the time averaged quadratic operator 
 $ \overline{T} = \int_{-\infty}^{\infty}\mathopen{:}\dot{\phi}^{\,2}(t,0)\mathclose{:}f(t)dt$, where $\phi$ is a massless minimally coupled scalar field and $f(t)$ is the sampling function . We use a 
 dimensionless variable $x \propto \overline{T}\tau^4$, where $\tau$ is a characteristic timescale of the sampling function. 
 Numerical results for both Lorentzian and compactly supported functions show that the 
 probability distribution of vacuum quantum fluctuations is bounded below at $x=-x_0 < 0$,  and that the tail of the probability distribution varies as an incomplete gamma function in agreement with the 
 previous studies~\cite{Fewster:2012ej, Fewster:2015hga}. We apply a best-fit procedure through a least-squares method to the tail of the $P_>(x)$-curves in order to determine
values for parameters in Eq.~(\ref{bestfit}). The results for $p_1, a, b, \text{and}\,c$ parameters show good agreement with the predictions of the high moments approach. (See 
Table~\ref{parametersalphas}.) The diagonalization procedure is able to reproduce with great accuracy the rate of decrease of the tail of the cumulative probability distribution. We reproduce the 
relation $c = \alpha/3$ for $\alpha= \left( 0.5,0.6,0.7,1 \right)$, where $\alpha=1$ corresponds to the case of the Lorentzian function, with percentage errors  less than $6\%$ compared to the theoretical 
values predicted by the high moments approach~\cite{Fewster:2012ej, Fewster:2015hga}. Our results confirm that averaging over a finite time 
interval, with compactly supported functions, results in a  probability distribution which falls more slowly than for the case of Lorentzian averaging, and both fall more slowly than 
exponentially.    

Recall that we have quantized the scalar field in a sphere with finite radius $R$, so the probability distribution which we calculate could differ from that of empty Minkowski spacetime.
As was noted in Sect.~\ref{CDF}, there should be no difference for the $\alpha = 1/2$ case, as the duration of the sampling is less than the light travel time to the boundary and back. In the other
cases, there could in principle be an effect of the boundary. However, this is likely only to alter the lower frequency modes, which are not expected to give a large contribution to the
tail of the distribution. 

The diagonalization method is free of the ambiguity potentially present in the high moments approach, and leads to a unique result for the probability distribution. It also has the potential to determine 
the entire distribution, including its lower bound, which is also the optimum quantum inequality bound on expectation values of the averaged operators.  In addition, it can provide information
about the particle content of the eigenstates of the averaged stress tensor which are associated with the large fluctuations.                

\section{acknowledgments}
We would like to thank Tom Roman for valuable discussions in the early stage of this project. CJF thanks Simon Eveson for a useful discussion concerning the numerical evaluation of~\eqref{eq:Cshift_theoretical}. This work was supported in part by the National Science Foundation under Grant PHY-1607118.

\appendix
\section{} 
\label{appendixA}

The expression $\Lambda = \sqrt{U^{T}K(F+G)K^TU}$ entails $U\Lambda^2U^{-1}=K(F+G)K^T$, where 
$K^{\dagger}=K^T$ and $U^{-1}=U^{T}$. Then, with $\Phi = K^TU\Lambda^{-1/2}$ and $\Psi = (F+G)\Phi\Lambda^{-1}$, we have
\begin{equation}\label{FplusG}
(F+G)\Phi =\Psi \Lambda
\end{equation}
by definition and also $K\Psi = K(F+G)\Phi\Lambda^{-1}= U\Lambda^{1/2}$. Then 
\begin{equation}
(F-G)\Psi = K^TK\Psi = K^T U\Lambda^{1/2}=\Phi \Lambda\,.
\label{FminusG}
\end{equation}
Using the definitions of $A$ and $B$ from Eq.~(\ref{AandB}), Equations~(\ref{FplusG}) and~(\ref{FminusG}) lead to Eqs.~(\ref{FGplus1}) and~(\ref{FGminus1}), respectively, according to
\begin{align}
(F+G)(A+B)=(F+G)\Phi=\Psi \Lambda = (A-B)\Lambda\,,\\
(F-G)(A-B)=(F-G)\Psi=\Phi \Lambda = (A+B)\Lambda\,.
\end{align}
Finally, using $\Phi\Psi^T = K^TU\Lambda^{-1/2}(K^{-1}U\Lambda^{1/2})^T=I$ and hence $\Psi\Phi^T = I$, we have
\begin{align}
AA^T - BB^T &= \frac{1}{2}(\Phi \Psi^T + \Psi \Phi^T)=I\,,\\
AB^T - BA^T &= \frac{1}{2}(\Psi\Phi^T - \Phi \Psi^T)=\frac{1}{2}(I-I)=0\,,
\end{align}
where $I$ and $0$ correspond to the identity and null matrices, respectively. These equations are the conditions that $A$ and $B$ have to satisfy in order to define a Bogoliubov transformation, 
Eq.~(\ref{condition}).

\section{}
\label{appendixB}
We listed below probabilities of finding specific outcomes
 in a measurement of a time averaged normal ordered quadratic operator. We have only considered up to 
 the 6-particle sector taking out the outcomes given by Eqs.~(\ref{TakeoutOutcome1}) and (\ref{TakeoutOutcome2}).
 It is understood that the coefficients of the $\mathcal{M}$ matrix, which appear in Table~\ref{TablePxOutcomes}, come from the diagonalization procedure explained in Sec.~\ref{diagonalization}. 
 \begin{table}[ht]
\begin{threeparttable}
\caption{Probabilities and outcomes of a time averaged normal ordered quadratic operator.} 
\label{TablePxOutcomes} 
\centering 
\begin{tabular}{ c | c} 
\hline\hline 
{\bf{ Probability}} & {\bf{Outcome}} \\ [0.2ex] 
\hline 
 $|\mathcal{N}|^2$ & $C_{\text{shift}}$  \\ 
 \hline 
 $|\mathcal{N}|^2|\mathcal{M}_{ij}|^2$ & $\lambda_i + \lambda_j + C_{\text{shift}}^{\hspace{8 mm}(a)}$  \\
 \hline 
  $\frac{1}{2}|\mathcal{N}|^2|\mathcal{M}_{ii}|^2$ & $2\lambda_i + C_{\text{shift}}$  \\
  \hline 
 $\frac{3}{8}|\mathcal{N}|^2|M_{ii}|^4$ & $4\lambda_i + C_{\text{shift}}$   \\
 \hline 
  $|\mathcal{N}|^2|\mathcal{M}_{ij}^2 + \frac{1}{2}\mathcal{M}_{ii}\mathcal{M}_{jj}|^2$ & $2\lambda_i + 2\lambda_j +  C_{\text{shift}}^{\hspace{8 mm}(a)}$  \\
  \hline 
  $\frac{1}{2}|\mathcal{N}|^2|\mathcal{M}_{ii}\mathcal{M}_{jk}+2\mathcal{M}_{ij}\mathcal{M}_{ik}|^2$ & $2\lambda_i + \lambda_j + \lambda_k + C_{\text{shift}}^{\hspace{8 mm}(b)}$   \\
  \hline 
  $\frac{3}{2}|\mathcal{N}|^2|\mathcal{M}_{ii}|^2|\mathcal{M}_{ij}|^2$ & $3\lambda_i + \lambda_j +  C_{\text{shift}}$ \\
  \hline 
  $|\mathcal{N}|^2|\mathcal{M}_{il}\mathcal{M}_{jk}+\mathcal{M}_{ik}\mathcal{M}_{jl}+\mathcal{M}_{ij}\mathcal{M}_{kl}|^2$ & $\lambda_i +
   \lambda_j +  \lambda_k + \lambda_l + C_{\text{shift}}^{\hspace{8 mm}(c)}$ \\
  \hline  
 $\frac{5}{16}|\mathcal{N}|^2|\mathcal{M}_{ii}|^6$ & $6\lambda_i + C_{\text{shift}}$  \\
 \hline 
    $\frac{15}{8}|\mathcal{N}|^2|M_{ii}|^4|\mathcal{M}_{ij}|^2$ & $5\lambda_i + \lambda_j + C_{\text{shift}}$  \\
    \hline 
    $\frac{3}{16}|\mathcal{N}|^2| 4\mathcal{M}_{ii}\mathcal{M}_{ij}^2+\mathcal{M}_{ii}^2\mathcal{M}_{jj} |^2$ & $4\lambda_i + 2\lambda_j + C_{\text{shift}}$ \\
    \hline 
     $\frac{1}{4}|\mathcal{N}|^2| 2\mathcal{M}_{ij}^3+3\mathcal{M}_{ii}\mathcal{M}_{ij}\mathcal{M}_{jj} |^2$ & $3\lambda_i + 3\lambda_j + C_{\text{shift}}^{\hspace{8 mm}(a)}$\\
     \hline 
     $\frac{1}{8}|\mathcal{N}|^2| 2\mathcal{M}_{ik}^2\mathcal{M}_{jj}+8\mathcal{M}_{ij}\mathcal{M}_{ik}\mathcal{M}_{jk}+$ & $2\lambda_i + 2\lambda_j + 
     2\lambda_k + C_{\text{shift}}^{\hspace{8 mm}(d)}$\\
    $2\mathcal{M}_{ii}\mathcal{M}_{jk}^2+2\mathcal{M}_{ij}^2\mathcal{M}_{kk}+\mathcal{M}_{ii}\mathcal{M}_{jj}\mathcal{M}_{kk}|^2$ &\\
    \hline 
    $\frac{3}{8}|\mathcal{N}|^2|\mathcal{M}_{ii}^2\mathcal{M}_{jk}+4\mathcal{M}_{ii}\mathcal{M}_{ij}\mathcal{M}_{ik}|^2$ & $4\lambda_i + \lambda_j + \lambda_k + C_{\text{shift}}^{\hspace{8 mm}(b)}$\\
    \hline 
       $\frac{1}{4}|\mathcal{N}|^2| 2\mathcal{M}_{ik}\mathcal{M}_{il}\mathcal{M}_{jj}+$ & $2\lambda_i + 2\lambda_j + \lambda_k + \lambda_l + C_{\text{shift}}^{\hspace{8 mm}(e)}$\\
      $4\mathcal{M}_{ij}\mathcal{M}_{il}\mathcal{M}_{jk} +4\mathcal{M}_{ij}\mathcal{M}_{ik}\mathcal{M}_{jl}+$ &  \\
       $2\mathcal{M}_{ii}\mathcal{M}_{jk}\mathcal{M}_{jl}+2\mathcal{M}_{ij}^2\mathcal{M}_{kl}+\mathcal{M}_{ii}\mathcal{M}_{jj}\mathcal{M}_{kl}|^2$\\
       \hline 
       $\frac{3}{4}|\mathcal{N}|^2| 2\mathcal{M}_{ij}^2\mathcal{M}_{ik}+$ & $3\lambda_i + 2\lambda_j + \lambda_k + C_{\text{shift}}^{\hspace{8 mm}}$\\
       $\mathcal{M}_{ii}\mathcal{M}_{ik}\mathcal{M}_{jj}+2\mathcal{M}_{ii}\mathcal{M}_{ij}\mathcal{M}_{jk}|^2$ & \\
       \hline 
       $\frac{3}{2}|\mathcal{N}|^2|2\mathcal{M}_{ij}\mathcal{M}_{ik}\mathcal{M}_{il}+\mathcal{M}_{ii}\mathcal{M}_{il}\mathcal{M}_{jk}+$ & $3\lambda_i + \lambda_j + \lambda_k + 
       \lambda_l + C_{\text{shift}}^{\hspace{8 mm}(f)}$\\
       $\mathcal{M}_{ii}\mathcal{M}_{ik}\mathcal{M}_{jl}+\mathcal{M}_{ii}\mathcal{M}_{ij}\mathcal{M}_{kl}|^2$ & \\
\hline\hline 
\end{tabular}
    \begin{tablenotes}
     \scriptsize
      \item Here we have $^{(a)}(i<j)\,,\,\,^{(b)}(j<k)\,,\,\,^{(c)}(i<j<k<l)\,,\,\,^{(d)}(i<j<k)\,,\,\,^{(e)}(i<j) \cap (k<l)\,,\,\,^{(f)}(j<k<l)\,,$ and the $\mathcal{M}$ matrix is defined in Sec.~\ref{multimodecase}.
    \end{tablenotes}
    \end{threeparttable}
\end{table} 



\begin{thebibliography}{00}
\bibitem{MCT}
To obtain the corresponding eigenvalues of the dynamical matrix, we use the Multiprecision Computing Toolbox for MATLAB 4.4.3.12625 
developed by Advanpix LLC., Yokohama, Japan.,
\end{thebibliography}
\end{document}